\documentclass[aps,pra,twocolumn,superscriptaddress,amsmath]{revtex4-2}	
\usepackage{graphicx}       
\usepackage{color}
\usepackage{hyperref}
\usepackage{cancel}
  
\graphicspath{{figures_manuscript/}}
\usepackage{verbatim} 
\usepackage[normalem]{ulem}      
 
\begin{document} 
 
\title{Quantum interference in the resonance fluorescence of a 
$J=1/2 - J'=1/2$ \\ atomic system: 
Quantum beats, nonclassicality, and non-Gaussianity }
 
\author{H. M. Castro-Beltr\'an} 
\email{hcastro@uaem.mx}
\affiliation{Centro de Investigaci\'on en Ingenier\'ia y Ciencias 
Aplicadas and Instituto de Investigaci\'on en Ciencias B\'asicas y Aplicadas, \\ 
Universidad Aut\'onoma del Estado de Morelos, 
Avenida Universidad 1001, 62209 Cuernavaca, Morelos, M\'exico}
\author{O. de los Santos-S\'anchez}
\affiliation{Tecnologico de Monterrey, Escuela de Ingenier\'ia y Ciencias,
Ave. Carlos Lazo 100, Santa Fe, Mexico City, M\'exico, 01389}
\author{L. Guti\'errez}
\affiliation{Instituto de Ciencias F\'isicas,  Universidad Nacional Aut\'onoma 
de M\'exico, \\ 62210 Cuernavaca, Morelos, M\'exico}
\author{A. D. Alcantar-Vidal}
\affiliation{Centro de Investigaci\'on en Ingenier\'ia y Ciencias 
Aplicadas and Instituto de Investigaci\'on en Ciencias B\'asicas y Aplicadas, \\ 
Universidad Aut\'onoma del Estado de Morelos, 
Avenida Universidad 1001, 62209 Cuernavaca, Morelos, M\'exico}
 
\date{\today}
\begin{abstract}
We study theoretically quantum statistical and spectral properties of the 
resonance fluorescence of a single atom or system with angular 
momentum $J=1/2 - J'=1/2$ driven by a monochromatic linearly polarized 
laser field, due to quantum interference among its two antiparallel, $\pi$ 
transitions. A magnetic field parallel to the laser polarization is applied to 
break the degeneracy (Zeeman effect). In the nondegenerate case, the 
$\pi$ transitions evolve at different generalized Rabi frequencies, producing  
quantum beats in the intensity and the dipole-dipole, intensity-intensity, and 
quadrature-intensity correlations. For a strong laser and large Zeeman 
splitting the beats have mean and modulation frequencies given by the 
average and difference, respectively, of the Rabi frequencies, unlike the
beats studied in many spectroscopic systems, characterized by a 
modulated exponential-like decay. Further, the Rabi frequencies are those 
of the pairs of sidebands of the Mollow-like spectrum of the system. In the 
two-time correlations, the cross contributions, i.e., those with products of 
probability \textit{amplitudes} of the two $\pi$ transitions, have a lesser role 
than those from the interference of the probability densities. In contrast, there 
are no cross terms in the total intensity.We also consider nonclassical and 
non-Gaussian properties of the phase-dependent fluorescence for the cases 
of weak to moderate excitation and in the regime of beats. The fluorescence 
in the beats regime is nonclassical, mainly from third-order dipole fluctuations, 
which reveal them to be also strongly non-Gaussian, and their quadrature 
spectra show complex features around the Rabi frequencies. For small laser 
and Zeeman detunings, a weak to moderate laser field pumps the system 
partially to one of the ground states, showing slow decay in the two-time 
correlations and a narrow peak in the quadrature spectra.    
\end{abstract}

\maketitle

 \section{Introduction} 
In spectroscopy, quantum beats are the modulation of the spontaneous 
emission decay of a multilevel system due to the energy difference among 
its excited or its ground states, the beats resulting from the indeterminacy 
of a photon's path when observed by a broadband detector. Two-level 
systems with degenerate upper and ground levels can have their levels 
split with a magnetic field and give way to quantum beats. In recent weak 
field cavity QED experiments beats were observed from spontaneous 
emissions to the ground states \cite{NOBC10,CSP+13}. As is often the case 
of beats using ground states, a higher-order process may be required to 
establish the coherence among them and then observe beats in intensity 
correlations \cite{Zajonc}. The variety of schemes and conditions to produce 
quantum beats is too big to consider here, see 
Ref.\cite{Lee+23,WuLiWu23,FiSw05}. 

Resonance fluorescence, the continued spontaneous emission by atoms 
under laser excitation, is a well-established means in quantum optics for the 
study of, among others: nonclassical properties of the field \cite{Carm02}, 
such as antibunching and squeezing; atomic coherence \cite{FiSw05}, such 
as electromagnetically induced transparency; and in atomic physics with laser 
cooling and trapping \cite{Metcalf} and atomic structure \cite{Budker}, such 
as determination of transition frequencies and level shifts. 

Recently, the properties of the resonance fluorescence of a single atomic 
system with angular momentum transition $J=1/2 - J'=1/2$ driven by a 
monochromatic laser have been the subject of great interest due to the 
possibility of observing vacuum-induced coherence effects due to 
interference among the two antiparallel $\pi$ transitions, emitting into the 
same frequency range of the electromagnetic vacuum. Here, the $\pi$ 
transitions are incoherently coupled, mediated by spontaneous emission in 
the $\sigma$ transitions, and excited again by the laser. Being the dipoles of 
the transitions antiparallel it is now realistic to observe interference effects, 
while the $V$ and $\Lambda$ three-level systems require additional 
preparation because the transitions are perpendicular \cite{FiSw05,FiSw04}. 
Particular attention has been devoted to the spectrum 
\cite{PoSc76,KiEK06a,KiEK06b,KMEK10}, time-energy complementarity 
\cite{KiEK06a,KiEK06b}, Young's interference \cite{EBB+92}, phase shifts 
\cite{FSA+17}, nonclassical properties of the fluorescence via squeezing 
\cite{TaXL09}, photon correlations \cite{DasAg08}, frequency-resolved 
photon correlations \cite{ZWLi20}, and cooperative effects in photon 
correlations \cite{WRZS20}. Also, the case of additional laser excitation of 
one of the $\sigma$ transitions on the spectrum and squeezing has been 
studied in \cite{Bergou99,Arun19,Arun20}. Curiously, in those works, there 
is no mention of quantum beats, which are among the more familiar 
manifestations of quantum interference. 

In this paper, we investigate theoretically quantum interference effects, 
nonclassicality, and non-Gaussianity of the resonance fluorescence light 
from the $\pi$ transitions of a single $J=1/2 - J'=1/2$ atomic system driven 
by a linearly polarized laser and a magnetic field to break the degeneracy. 
Both the unequal Zeeman splittings for ground and excited levels and the 
laser-induced ac Stark effect makes the effective transition frequencies 
different and also the $\pi$ transitions evolve with different generalized 
Rabi frequencies $\Omega_1$ and $\Omega_2$; these determine 
dynamical and spectral properties. Resonance fluorescence in this system 
then goes beyond pure spontaneous emission, and the generalized Rabi 
frequencies replace energy-level splittings. For a start, we illustrate the 
\textit{evolution} of the state populations, an aspect bypassed previously in 
favor of their steady-state values, which helps us to understand the origin of 
interference in this system. 

As a major result, we find the emergence of quantum beats due to the 
interference of light from the $\pi$ transitions in the nondegenerate situation.  
For weak laser and magnetic fields $\Omega_1$ and $\Omega_2$ are 
nearly equal, so the beats have simple damped oscillatory behavior or just 
damped decay. For strong laser and magnetic fields, on the other hand, the 
generalized Rabi frequencies split enough such that the beats evolve to 
damped wave packets with well-defined oscillations at the \textit{average} 
frequency $(\Omega_1 + \Omega_2)/2$ and \textit{modulation} frequency 
$(\Omega_2 - \Omega_1)/2$. There are no beats in the degenerate case. 

We study quantum beats in the fluorescence intensity, where an initial 
condition with unequal ground state populations is needed, and in two-time 
correlations such as dipole-dipole (to calculate spectra), intensity-intensity 
\cite{HBT,Glauber-g2}, and phase-dependent intensity-amplitude 
correlations \cite{CCFO00,FOCC00}. For the two-time 
correlations it is only necessary that the dominant terms in the correlation be 
initially nonzero. An early hint for the appearance of beats may be seen in 
the closeness of sidebands in the spectrum at the frequencies 
$\pm \Omega_1$ and $\pm \Omega_2$ \cite{KiEK06b}; the corresponding 
dipole correlation in the Wiener-Khinchine formula exhibits strongly 
damped quantum beats. 

\begin{figure}[t]
\includegraphics[width=7.5cm,height=6.0cm]{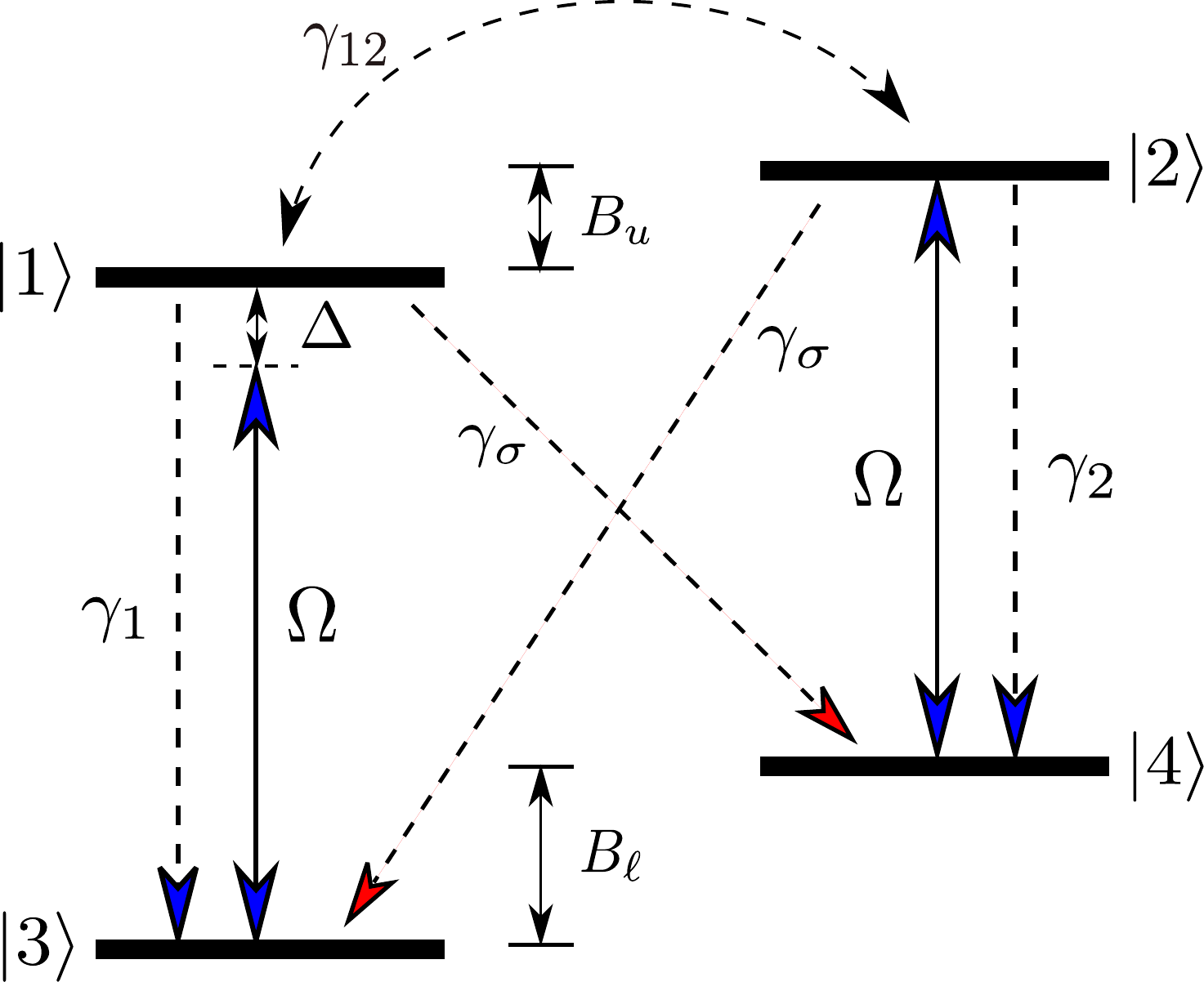} 
\caption{\label{fig:4LA} 
Scheme of the $J=1/2$ -- $J=1/2$ atomic system interacting with 
a laser driving the $|1 \rangle - |3 \rangle$ and $|2 \rangle - |4 \rangle$ 
transitions with Rabi frequency $\Omega$ and detuning $\Delta$. 
There are spontaneous decay rates $\gamma_1$, $\gamma_2$ and 
$\gamma_{\sigma}$, vacuum-induced coherence $\gamma_{12}$, 
and Zeeman frequency splittings $B_{\ell}$ and $B_u$.  }
\end{figure}

While beats may be obtained from incoherent sources \cite{Forrester} 
some form of coherence is needed. One may ask if the vacuum-induced 
coherence (VIC), the coupling of the upper states (in this case) by the 
reservoir, is such. In this paper, the incoherent sources are the $\pi$ 
transitions emitting photons spontaneously. These transitions are 
incoherently coupled by the spontaneous emission in the $\sigma$ 
transitions. Coherence is maintained because all emissions make the atom 
end up in one ground state or the other, i.e., the system is closed, see 
Fig.~\ref{fig:4LA}. We find that the cross terms in the correlations that link 
both $\pi$ transitions, the signature of VIC, are small compared to those of 
interference from the single transitions. This is also observed in the 
contributions to the total intensity from the coherent and incoherent parts. 

The two-time amplitude-intensity correlation (AIC) is a measurement that 
probes up to third-order quadrature fluctuations that may violate classical 
inequalities \cite{CCFO00,FOCC00}; single few-level atom resonance 
fluorescence does \cite{CaRG16,GCRH17,SaCa21,ScVo05,ScVo06}. 
In the weak-field regime, the AIC is reduced to a second-order detection, 
i. e., of squeezed light \cite{Carm02}, and the fluctuations are Gaussian. 
For stronger laser fields, squeezing, if any, is mixed with third-order 
fluctuations. In this nonlinear interaction case, with quantum beats as a 
notable feature, the light's fluctuations are also non-Gaussian. The AIC 
becomes a tool for studying non-Gaussianity of light, in both the time and 
frequency domains.  

We work around two regimes in the parameter space: (i) the Rabi frequency,  
laser detuning and the difference Zeeman splittings are close to the 
spontaneous emission rate, enough to exhibit beats, but there is some 
squeezing, and (ii) where both laser and magnetic fields are strong, with 
well-developed beats that contribute to large third-order nonclassical and 
non-Gaussian fluctuations. 
  
Finally, we also analyze squeezing via the variance or total noise 
\cite{WaMi08}, finding that the total squeezing is limited to a region in 
parameter space of small laser and Zeeman detunings, and the laser 
intensity should be even smaller than in the two-level atom case. Beats are 
usually well outside the squeezing region. 
  
The paper is organized as follows. After describing the model's main features 
in Sec. II, we discuss the basic dynamic and stationary properties 
of the atomic expectation values in Sec. III. Then, in Sec. IV, we 
describe the scattered field intensity and its fluctuations. Secs. V-VII are 
devoted to two-time correlation functions. In Sec. V, we relate the 
double sideband spectrum \cite{KiEK06b} with beats in the dipole-dipole 
correlation function. Then, in Sec. VI, we study Brown-Twiss 
photon-photon correlations \cite{HBT,Glauber-g2}, extending the work of 
Ref.\cite{DasAg08} to the nondegenerate case. Section VII is devoted to  
phase-dependent fluctuations by conditional homodyne detection  
\cite{CCFO00,FOCC00} in both the temporal and spectral domains. In 
Sec. VIII, we consider squeezing of fluctuations using the variance. 
We close in Sec. IX with a discussion and conclusions. Two Appendices 
detail solution methods, initial conditions, and optimal appearance of beats.

\section{Model} 
The system, illustrated in Fig.~\ref{fig:4LA}, consists of a two-level atom 
with transition $J=1/2$  -- $J=1/2$ and states with magnetic quantum 
number $m=\pm J$,   
\begin{eqnarray}
|1\rangle &=& |J, -1/2 \rangle , 	\qquad |2\rangle = |J, 1/2 \rangle , 	
	\nonumber \\ 
|3\rangle &=& |J, -1/2 \rangle , 	\qquad  |4\rangle = |J, 1/2 \rangle .
\end{eqnarray}
The matrix elements are
\begin{eqnarray} 	\label{eq:dipoles}
\mathbf{d}_1 &=& \langle 1| \hat{\mathbf{d}} |3 \rangle 
	= - \frac{1}{\sqrt{3}} \mathcal{D} \mathbf{e}_z , 	\qquad 
	\mathbf{d}_2 = \langle 2| \hat{\mathbf{d}} |4 \rangle  = - \mathbf{d}_1 , 
	\nonumber \\ 
\mathbf{d}_3 &=& \langle 2| \hat{\mathbf{d}} |3 \rangle 
	=  \sqrt{ \frac{2}{3} } \mathcal{D} \mathbf{e}_- , 	\qquad 
	\mathbf{d}_4 = \langle 1| \hat{\mathbf{d}} |4 \rangle  = \mathbf{d}_3^\ast , 
\end{eqnarray}
where $\mathcal{D}$ is the reduced dipole matrix element.  We choose 
the field polarization basis $\{\mathbf{e}_z, \mathbf{e}_-, \mathbf{e}_+\}$ 
(linear, left circular, right circular), where 
$\mathbf{e}_{\pm} = \mp (\mathbf{e}_x \pm i \mathbf{e}_y)/2$. 

The $\pi$ transitions, $|1\rangle -|3\rangle$ and $|2\rangle -|4\rangle$ 
($m=m'$), are coupled to linearly polarized light and have their dipole 
moments antiparallel. On the other hand, the $\sigma$ transitions,  
$|1\rangle -|4\rangle$ and $|2\rangle -|3\rangle$ ($m \neq m'$), are 
coupled to circularly polarized light. This configuration can be found, 
for example, in $^{198} \mathrm{Hg}^+$  \cite{PoSc76}, and 
$^{40} \mathrm{Ca}^+$ \cite{WRZS20}. 

The level degeneracy is removed by the Zeeman effect, applying a 
magnetic field $B_z$ along the $z$ direction. Note that the energy splittings 
$g \mu_B B_z$ of the upper ($u$) and lower ($\ell$) levels are different due 
to unequal Land\'e $g$ factors, $g_u$ and $g_{\ell}$, respectively; $\mu_B$ 
is Bohr's magneton. The difference Zeeman splitting is 
\begin{equation} 	\label{eq:delta}
\delta = \frac{(g_u -g_{\ell}) \mu_B B_z}{\hbar} 
	= \frac{g_u -g_{\ell}}{g_{\ell}}  B_{\ell} , 
\end{equation}
where $B_{\ell} = g_l \mu_B B_z /\hbar$. For $^{198} \mathrm{Hg}^+$ 
$g_u =2/3$ and $g_{\ell} = 2$, so 
$\hbar \delta = -(4/3) \mu_B B_z = -(2/3) \hbar B_{\ell}$. 

The atom is driven by a monochromatic laser of frequency $\omega_L$, 
linearly polarized in the $z$ direction, propagating in the $x$ direction, 
\begin{eqnarray} 	\label{eq:laser}
\mathbf{E}_L (x,t) &=& E_0 e^{i (\omega_L t - k_L x)} \mathbf{e}_z 
	+\mathrm{c.c.}  ,
\end{eqnarray}
thus driving only the $\pi$ transitions.  

The free atomic, $H_0$, and interaction, $V$, parts of the Hamiltonian 
are, respectively: 
\begin{eqnarray}
H_0 &=& \hbar \omega_{13} A_{11} +  \hbar (\omega_{24} +B_{\ell}) A_{22} 
	+ \hbar B_{\ell} A_{44} ,	\\ 
V &=& \hbar \Omega (A_{13} -A_{24})  e^{i \omega_L t} + \mathrm{h.c.}
\end{eqnarray}
where $A_{jk} = |j \rangle \langle k|$ are atomic operators, $\omega_{13}$ 
and $\omega_{24} = \omega_{13} +\delta$ are the frequencies of the 
$|1 \rangle - |3 \rangle$ and $|2 \rangle - |4 \rangle$ transitions, 
respectively, and $\Omega = E_0 \mathcal{D} / \sqrt{3}\,\hbar$ is the 
Rabi frequency. The frequencies of the other transitions are 
$\omega_{23} =\omega_{13} -\delta$ and $\omega_{14} 
= \omega_{13} -B_{\ell}$. Using the unitary transformation
\begin{equation}
U = \exp{[ (A_{11} +A_{22}) i \omega_L t ]} , 
\end{equation}
the Hamiltonian in the frame rotating at the laser frequency is
\begin{eqnarray} 	\label{eq:Hamiltonian}
H &=& U^\dag (H_0 +V) U , 	\nonumber \\
&=& - \hbar \Delta A_{11} - \hbar (\Delta -\delta) A_{22}  
	+\hbar B_{\ell} (A_{22} +A_{44}) \nonumber	\\ 
&& + \hbar \Omega \left[ (A_{13} -A_{24})  + \mathrm{h.c.} \right] ,
\end{eqnarray}
where $\Delta = \omega_L - \omega_{13}$ is the detuning of the laser 
from the $|1 \rangle - |3 \rangle$ resonance transition, and 
$\Delta - \delta$ is the detuning on the $|2 \rangle - |4 \rangle$ transition. 

The excited states decay either in the $\pi$ transitions emitting photons 
with linear polarization at rates $\gamma_1 = \gamma_2$, or in the 
$\sigma$ transitions emitting photons of circular polarization at a rate 
$\gamma_{\sigma}$. There is also a cross-coupling of the excited states 
by the reservoir, responsible for part of the quantum interference we wish 
to study. In general, the decay rates are written as 
\begin{eqnarray}
\gamma_{ij} &=& \frac{\mathbf{d}_i \cdot \mathbf{d}_j^\ast}
	{|\mathbf{d}_i| |\mathbf{d}_j|} \sqrt{ \gamma_i \gamma_j } , 
	\qquad i,j=1,2 . 
\end{eqnarray}
In particular, we have $\gamma_{ii} = \gamma_1 = \gamma_2$ and 
$\gamma_{13} = \gamma_{24} = \gamma_{\sigma}$. Also, given that 
$\mathbf{d}_1$ and $\mathbf{d}_2$ are antiparallel, $\gamma_{12} 
= \gamma_{21} = - \sqrt{ \gamma_1 \gamma_2 } = -\gamma_1$. 

The total decay rate is 
\begin{eqnarray} 	\label{eq:totaldecay}
\gamma = \gamma_1 + \gamma_{\sigma} = \gamma_2 + \gamma_{\sigma} .
\end{eqnarray}
The decays for the $\pi$ and $\sigma$ transitions occur with the branching 
fractions $b_{\pi}$ and $b_{\sigma}$ \cite{KiEK06b}, respectively, 
\begin{subequations}
\begin{eqnarray}
\gamma_1 = \gamma_2 = b_{\pi} \gamma ,     \qquad b_{\pi} = 1/3 , \\
\gamma_{\sigma} = b_{\sigma} \gamma , 	\qquad b_{\sigma} = 2/3 . 
\end{eqnarray}
\end{subequations}

\section{Master Equation} 
The dynamics of the atom-laser-reservoir system is described by the 
master equation for the reduced atomic density operator, $\rho$. In a frame 
rotating at the laser frequency ($\tilde{\rho} = U \rho U^\dag$)  it is given by 
\begin{eqnarray} 	\label{eq:master}
\dot{\tilde{\rho}} 
 &=& -\frac{i}{\hbar} [H,\tilde{\rho}] +\mathcal{L}_{\gamma}  \tilde{\rho} , 
\end{eqnarray}
where $-(i/\hbar) [H,\tilde{\rho}]$ describes the coherent atom-laser interaction 
and $\mathcal{L}_{\gamma}  \tilde{\rho}$ describes the damping due to 
spontaneous emission \cite{KiEK06b,Agarwal74}. Defining 
\begin{eqnarray}
S_1^- &=& A_{31} , \quad 	S_2^- = A_{42} , \quad 
	S_3^- = A_{32} , \quad 	S_4^- = A_{41} , 	\nonumber\\
S_i^+ &=& (S_i^-)^\dag , 	 
\end{eqnarray}
the dissipative part is written as 
\begin{eqnarray} 	\label{eq:mastereq}
\mathcal{L}_{\gamma} \tilde{\rho} 
	&=& \frac{1}{2} \sum_{i,j=1}^2 \gamma_{ij} \left( 2S_i^- \tilde{\rho} S_j^+ 
	- S_i^+ S_j^- \tilde{\rho} -\tilde{\rho} S_i^+ S_j^-  \right) 	\nonumber \\ 
   && + \frac{\gamma_{\sigma}}{2} \sum_{i=3}^4 \left( 2S_i^- \tilde{\rho} S_i^+ 
	- S_i^+ S_i^- \tilde{\rho} -\tilde{\rho} S_i^+ S_i^-  \right) 	.
\end{eqnarray}

We now define the Bloch vector of the system as
\begin{eqnarray} 	\label{eq:BlochVector}
\mathbf{Q} &\equiv& \left( A_{11}, A_{12}, A_{13},  A_{14}, A_{21}, A_{22}, 
	A_{23}, A_{24}, 	\right. 	\nonumber \\ 
	&& \left. A_{31}, A_{32},  A_{33},  A_{34},  
	A_{41}, A_{42}, A_{43}, A_{44} \right)^T .  
\end{eqnarray}
The equations for the expectation values of the atomic operators, 
$\langle A_{jk} \rangle = \tilde{\rho}_{kj}$, are the so-called Bloch 
equations, which we write compactly as 
\begin{eqnarray} 	\label{eq:BlochEqs0}
\frac{d}{dt}  \langle \mathbf{Q}(t) \rangle 
	= \mathbf{M}_B \langle \mathbf{Q}(t) \rangle , 
\end{eqnarray}
where $\mathbf{M}_B$ is a matrix of coeficients of the full master equation, 
and the formal solution is  
\begin{eqnarray} 	
\langle \mathbf{Q}(t) \rangle 
	&=& e^{\mathbf{M}_B t} \langle \mathbf{Q}(0) \rangle 	.
\end{eqnarray}

Since in this paper we are interested only in properties of the fluorescence 
emitted in the $\pi$ transitions we use the simplifying fact, already noticed 
in \cite{DasAg08}, that the Bloch equations can be reduced to a 
homogeneous set of equations for the populations and the coherences of 
the coherently driven $\pi$ transitions only: 
\begin{eqnarray} 	\label{eq:BlochEqs1}
\langle \dot{A}_{11} \rangle &=& - \gamma \langle A_{11} \rangle 
	+i \Omega (\langle A_{31} \rangle -\langle A_{13} \rangle) 	, 
	\nonumber \\ 
\langle \dot{A}_{13} \rangle &=& -\left( \frac{\gamma}{2} 
	+i\Delta  \right) \langle A_{13} \rangle - i \Omega (\langle A_{11} \rangle 
	- \langle A_{33} \rangle) 	,  \nonumber \\
\langle \dot{A}_{22} \rangle &=& -\gamma \langle A_{22} \rangle 
	- i \Omega (\langle A_{42} \rangle - \langle A_{24} \rangle) , 
	\nonumber \\ 
\langle \dot{A}_{24} \rangle &=& -\left( \frac{\gamma}{2} 
	+i (\Delta -\delta) \right) \langle A_{24} \rangle 
	+ i \Omega (\langle A_{22} \rangle - \langle A_{44} \rangle)  , 
	\nonumber \\ 
\langle \dot{A}_{31} \rangle &=& -\left( \frac{\gamma}{2} 
	-i\Delta  \right) \langle A_{31} \rangle + i \Omega (\langle A_{11} \rangle 
	- \langle A_{33} \rangle) 	, \nonumber \\ 
\langle \dot{A}_{33} \rangle &=& \gamma_1 \langle A_{11} \rangle 
	+\gamma_{\sigma} \langle A_{22} \rangle 
	-i \Omega (\langle A_{31} \rangle - \langle A_{13} \rangle) 	, 
	\nonumber \\
\langle \dot{A}_{42} \rangle &=& -\left( \frac{\gamma}{2} 
	-i (\Delta -\delta) \right) \langle A_{42} \rangle 
	- i \Omega (\langle A_{22} \rangle - \langle A_{44} \rangle)   ,  
	\nonumber \\ 
\langle \dot{A}_{44} \rangle &=& \gamma_{\sigma}  \langle A_{11} \rangle    
	+\gamma_2 \langle A_{22} \rangle 
	+i \Omega (\langle A_{42} \rangle - \langle A_{24} \rangle) . 
\end{eqnarray}
We thus define a new Bloch vector,   
\begin{eqnarray} 	\label{eq:BlochVector1}
\mathbf{R} &\equiv& \left( A_{11}, A_{13}, A_{22}, A_{24}, 	
A_{31}, A_{33}, A_{42}, A_{44} \right)^T 
\end{eqnarray} 
and a corresponding matrix $\mathbf{M}$, Eq.~(\ref{eq:matrixM}), and 
solve as for $\langle \mathbf{Q}(t) \rangle$. Equations (\ref{eq:BlochEqs1}) 
depend on the magnetic field only through the difference Zeeman splitting, 
$\delta$, and do not depend on $\gamma_{12}$, the vacuum-induced 
coupling of the upper levels. This means that the coupling of the $\pi$ 
transitions occurs incoherently via spontaneous emission in the $\sigma$ 
transitions, yet they maintain coherence since the system is a closed one. 

The steady-state solutions, for which we introduce the short notation 
$\alpha_{jk} = \langle A_{jk} \rangle_{st}$, are  
\begin{subequations}
\begin{eqnarray} 	\label{eq:BlochSteady1}
\alpha_{11} &=& \alpha_{22} = \frac{\Omega^2}{2D} 	,\\ 
\alpha_{33} &=& \frac{ \Omega^2 +\gamma^2/4 +\Delta^2 }{2D}  ,\\ 
\alpha_{44} &=&  \frac{ \Omega^2 +\gamma^2/4 +(\Delta -\delta)^2 }	{2D } ,\\ 
\alpha_{13} &=& \frac{ \Omega (\Delta +i\gamma/2)}{2D} 	,\\ 
\alpha_{24} &=&  \frac{ \Omega (\delta -\Delta -i\gamma/2) }{2D}  ,\\ 
\alpha_{kj} &=& \alpha_{jk}^\ast . \nonumber 
\end{eqnarray}
\end{subequations} 
where 
\begin{equation} 	\label{eq:denominator}
D = 2\Omega^2 +\frac{\gamma^2 +\delta^2}{4}   
	+\left( \Delta -\frac{\delta}{2} \right)^2 	. 
\end{equation} 

Note also that in the degenerate system ($\delta =0$) 
$\alpha_{33} = \alpha_{44}$ and that $\alpha_{13} = - \alpha_{24}$,  
where the minus sign arises from the fact that the dipole moments 
$\mathbf{d}_1$ and $\mathbf{d}_2$ are antiparallel. 
\begin{figure}[t]
\includegraphics[width=8.5cm,height=6cm]{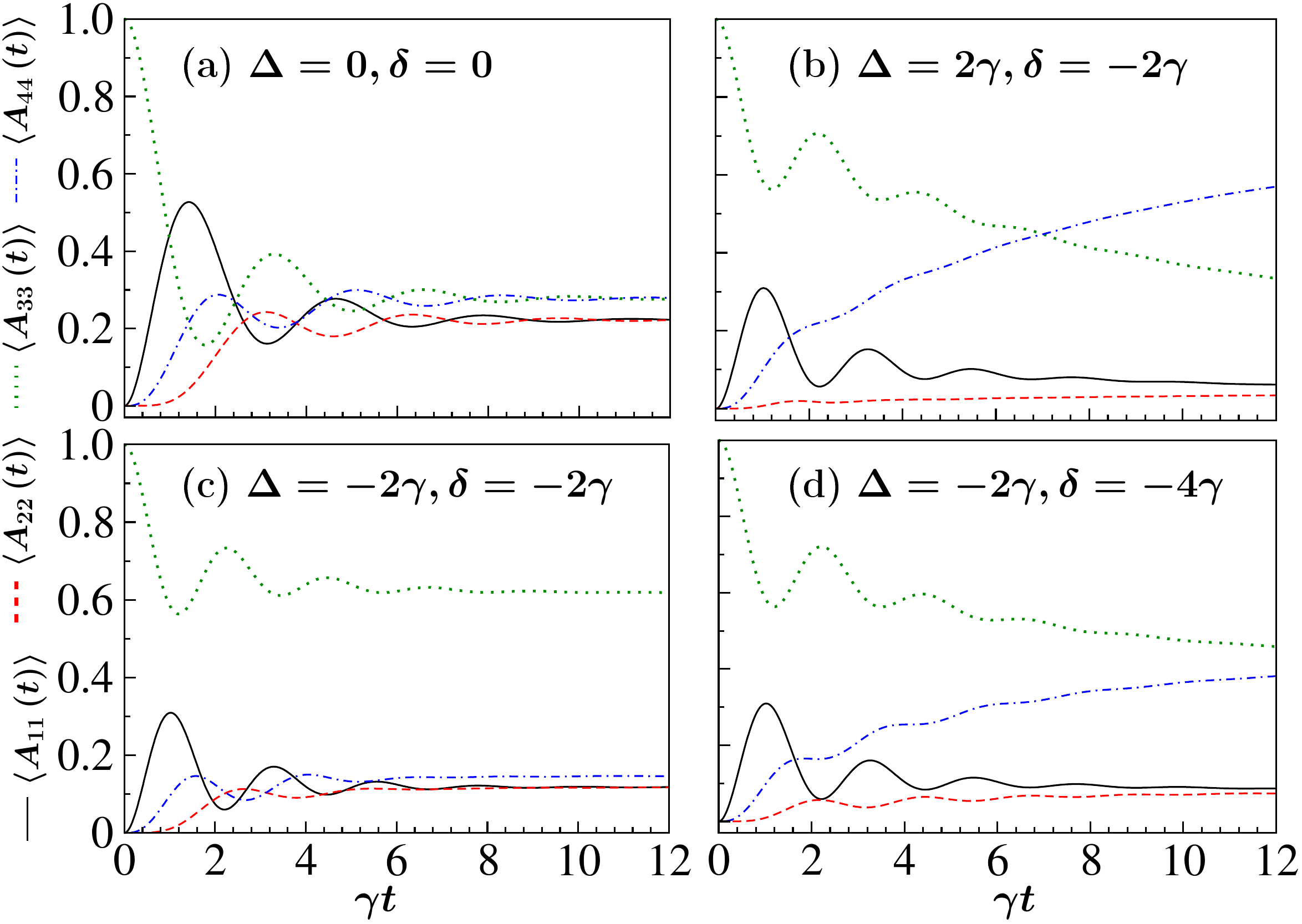} 
\caption{\label{fig:pops} 
Time-dependent populations $\langle A_{11}(t) \rangle$ (solid-black), 
$\langle A_{22}(t) \rangle$ (dashed-red), $\langle A_{33}(t) \rangle$ 
(dots-green), and $\langle A_{44}(t) \rangle$ (dashed-dots-blue), with 
the atom initially in state $|3 \rangle$. The parameters are: 
$\Omega=\gamma$ and (a) $\Delta = \delta=0$; 
(b) $\Delta = 2 \gamma$, $\delta= -2\gamma$; 
(c) $\Delta = \delta= -2\gamma$; 
(d) $\Delta =-2 \gamma$, $\delta= -4\gamma$.  }
\end{figure}

The remaining equations, which we omit, describe the evolution of the 
coherences of the $\sigma$ transitions 
($\langle A_{14} \rangle$, $\langle A_{23} \rangle$) and those among the 
upper and lower states  
($\langle A_{12} \rangle$, $\langle A_{34} \rangle$). These coherences 
vanish at all times because the dynamics of the $\sigma$ transitions is 
due only to spontaneous emission. Note, however, that for the calculation 
of fluorescence properties of the $\sigma$ transitions the complete set of 
Bloch equations, Eq.(\ref{eq:BlochEqs0}), is actually required. 

We gain valuable information on the nontrivial dynamics of the atomic 
system from the single-time expectation values bypassed in the previous 
literature on the system. In Fig.~\ref{fig:pops}, we show the populations for 
several particular cases, all with the atom initially in state $|3 \rangle$. In 
the degenerate case, $\delta=0$, the upper populations reach opposite 
phases by the end of the first Rabi cycle, Fig.~\ref{fig:pops}(a). This is 
understandable since the electron occupation of, say, state $|1\rangle$ 
implies not to be in state $|2\rangle$, and viceversa. A similar situation 
occurs for the lower populations. Next, we show three situations for the 
nondegenerate case with $\delta < 0$ (as it is for $^{198} \mathrm{Hg}^+$). 
In Fig.~\ref{fig:pops}(b) the laser is slightly detuned above the 
$|1\rangle - |3\rangle$ transition, but highly detuned from the 
$|2\rangle - |4\rangle$ transition, thus populating preferentially state 
$|4\rangle$ as optical pumping. In Fig.~\ref{fig:pops}(c) the laser is detuned 
below the $|1\rangle - |3\rangle$ transition, and the $|2\rangle - |4\rangle$ 
transition is now on resonance with the laser, but the population is pumped 
mainly to state $|3\rangle$. More visible in the latter case is that the excited 
state populations evolve with slightly different Rabi frequencies in the 
nondegenerate case. In Fig.~\ref{fig:pops}(d), we extend the previous 
case to a larger difference Zeeman splitting, such that the Rabi frequencies 
are equal, with the populations of the excited states again out of phase. In 
the next Sections, we will also consider the general cases of different Rabi 
frequencies and initial conditions with interesting consequences.  
\begin{figure}[t]
\includegraphics[width=8.5cm,height=6cm]{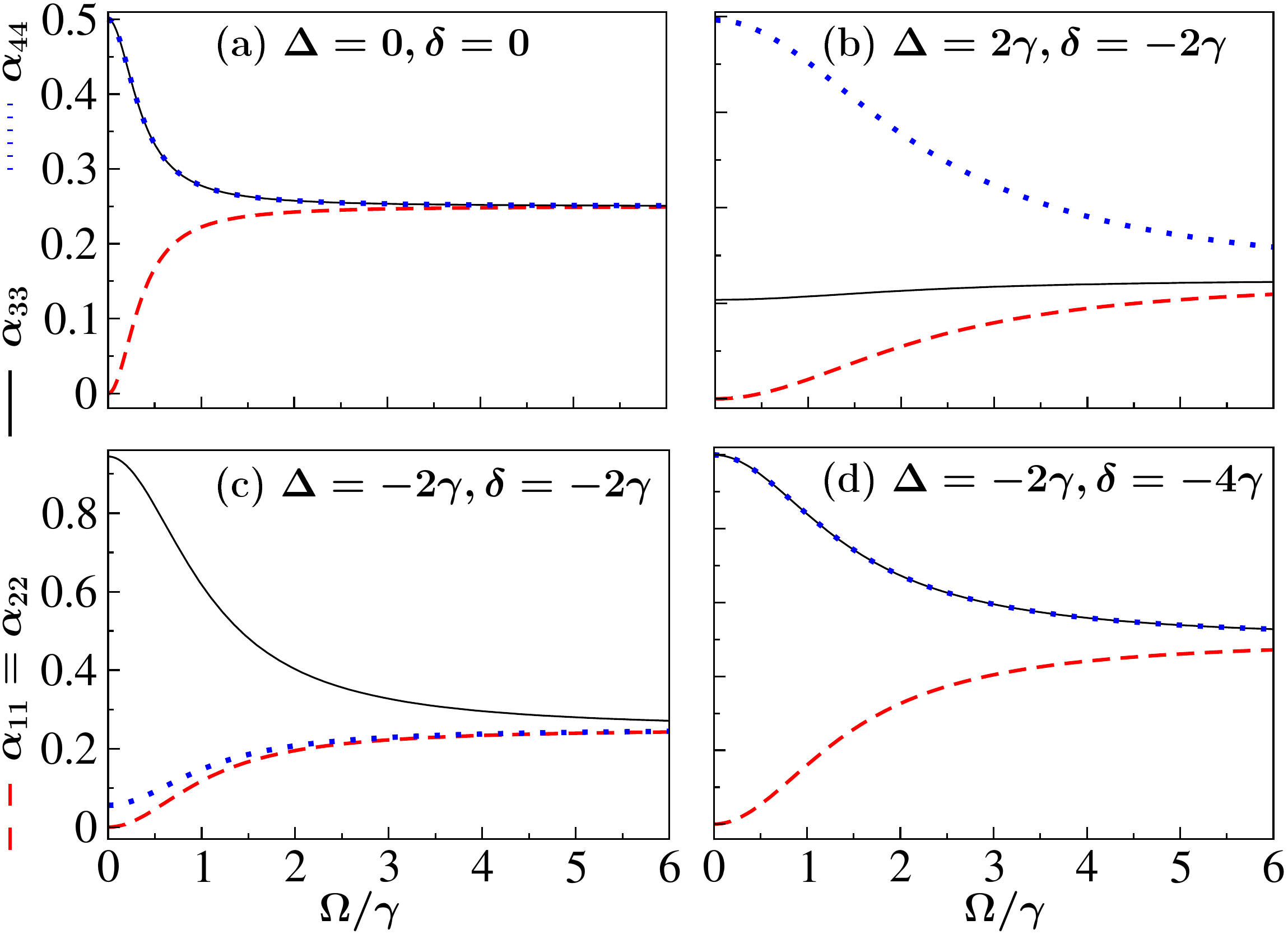} 
\caption{ \label{fig:pops_ss} 
Steady-state populations $\alpha_{jj}$ as a function of Rabi frequency: 
$\alpha_{11} = \alpha_{22}$ (dashed-red), $\alpha_{33}$ (solid-black), and 
$\alpha_{44}$ (dots-blue). All other parameters as in Fig.~\ref{fig:pops}.  }
\end{figure}

In Fig.~\ref{fig:pops_ss}, we show the steady-state populations as a function 
of the Rabi frequency; the other parameters are the same as in 
Fig.~\ref{fig:pops}. For strong fields, the populations tend to be equal (1/4), 
but arrive at that limit at different rates; for instance, for large detunings on 
both transitions, Fig.~\ref{fig:pops_ss}(d), it takes larger fields, as compared 
to the degenerate case, Fig.~\ref{fig:pops_ss}(a). On the other hand, for 
small detunings and weak to moderate fields, when one transition is closer to 
resonance than the other, the lower state of the more detuned transition is 
more populated, as seen in Figs.~\ref{fig:pops_ss} (b) and (c).

\section{The Scattered Field} 
In this Section, we study the main quantum dynamical and stationary 
properties of the intensity of the field scattered by the atom, with 
emphasis on the $\pi$ transitions. Although the intensity is not affected by 
interference, there are valuable insights analyzing the latter via intensity 
fluctuations. 

\subsection{Single-Time and Stationary Properties} 
In principle, the reservoir and laser radiation fields are quantized. However, 
for free-space atom-field interactions, it is customary to simplify the 
theoretical description by eliminating the reservoir modes, thus 
obtaining the master equation for the atomic system alone, 
Eq.~(\ref{eq:master}), and assuming the laser to be described by a classical 
quantity, Eq.~(\ref{eq:laser}), by a canonical transformation if the laser field 
is initially in a coherent state, and the reservoir in the vacuum state 
\cite{Mollow75}. 

But, then, it is also customary to keep the quantum nature of the scattered 
field because we are interested in the quantum fluctuations of the field 
emitted by the source. The atom, described by a dipole operator, absorbs 
photons from the laser and emits them to the reservoir. The scattering 
process is thus mediated by the reservoir, but now the focus is on the field, 
whereas in the master equation, the focus is on the atom. In a procedure 
with the dipole and rotating-wave approximations, a relevant reservoir 
frequency range and field initially in the vacuum state, the scattered field is 
proportional to the dipole operator, so the quantum nature of the atom 
leaves its imprint on the reemitted field, see, e.g., Ref.\cite{Carm02}.

The positive-frequency part of the emitted field operator is 
\cite{Agarwal74,Carm02} 
\begin{eqnarray}  	
\hat{E}^+ (\mathbf{r}, t) &=& \hat{E}_{\mathrm{free}}^+ (\mathbf{r}, t) 
	+\hat{E}_S^+ (\mathbf{r}, \hat{t}) ,
\end{eqnarray}
where $\hat{E}_{\mathrm{free}}^+ (\mathbf{r}, t)$ is the free-field part  
which, for a reservoir in the vacuum state, does not contribute to normally 
ordered correlations, hence we omit it in further calculations, and  
\begin{eqnarray} \label{eq:scatfield}
\hat{E}_S^+ (\mathbf{r}, t) &=& - \frac{\eta}{r} 
	\sum_{i=1}^4 \omega_i^2  \hat{\mathbf{r}} \times ( \hat{\mathbf{r}} 
	\times \mathbf{d}_i ) S_i^- (\hat{t})   
\end{eqnarray}
is the dipole source field operator in the far-field zone, where 
$\hat{t} = t -r/c$ is the retarded time and $\eta = (4\pi \epsilon_0 c^2)^{-1}$. 
Since $\omega_i \gg \gamma, \Omega,\Delta,\delta$, we may approximate 
the four transition as a single one $\omega_0$ in Eq.~(\ref{eq:scatfield}). 

Making $\hat{\mathbf{r}} = \mathbf{e}_y$ the direction of observation and 
using Eq.~(\ref{eq:dipoles}) we have 
\begin{eqnarray} 	\label{eq:ScattField}
\hat{E}_S^+ (\mathbf{r}, \hat{t}) 
&=& \hat{E}_{\pi}^+ (\mathbf{r}, \hat{t}) \,\mathbf{e}_z  
	+ \hat{E}_{\sigma}^+ (\mathbf{r}, \hat{t}) \,\mathbf{e}_x ,
\end{eqnarray}
i.e., the fields scattered from the $\pi$ and $\sigma$ transitions are 
polarized in the $\mathbf{e}_z$ and $\mathbf{e}_x$ directions, 
respectively, where 
\begin{subequations}
\begin{eqnarray}  	\label{eq:field-atomOps}
\hat{E}_{\pi}^+ (\mathbf{r}, \hat{t}) 
	&=&  f_{\pi}(r)   
	\left[  A_{31} (\hat{t})   -  A_{42} (\hat{t}) \right] , \\
\hat{E}_{\sigma}^+ (\mathbf{r}, \hat{t}) &=&  f_{\sigma}(r) 
	 \left[  A_{32} (\hat{t})  - A_{41} (\hat{t}) \right] , 
\end{eqnarray}
\end{subequations}
are the positive-frequency source field operators of the $\pi$ and 
$\sigma$ transitions, and 
\begin{eqnarray} 
f_{\pi}(r) = -\eta \omega_1^2 \mathcal{D}/\sqrt{3}r 	, \qquad 
	f_{\sigma}(r) = \sqrt{2} f_{\pi}(r) , 
\end{eqnarray}
are their geometric factors. 

The intensity in the $\pi$ transitions is given by 
\begin{subequations}
\begin{eqnarray} 	\label{eq:time_Intensity_pi}
I_{\pi} (\mathbf{r}, \hat{t}) &=& \langle \hat{E}_{\pi}^- (\mathbf{r}, \hat{t}) 
	\cdot 	\hat{E}_{\pi}^+ (\mathbf{r}, \hat{t}) \rangle 	\nonumber \\
&=& f_{\pi}^2(r) \langle A_{13} (\hat{t}) A_{31} (\hat{t}) 
	+A_{24} (\hat{t}) A_{42} (\hat{t}) \rangle  \nonumber \\
&=& f_{\pi}^2(r)  \langle A_{11} (\hat{t})  +A_{22} (\hat{t}) \rangle , 
\end{eqnarray} 
while in the steady state is
\begin{eqnarray} 	\label{eq:steady_intensity_pi}
I_{\pi}^{st}  &=& f_{\pi}^2(r) \left[ \alpha_{11} + \alpha_{22} \right] 
	=  \frac{\Omega^2}{D} . 
\end{eqnarray}
\end{subequations}
\begin{figure}[t]
\includegraphics[width=8.5cm,height=7cm]{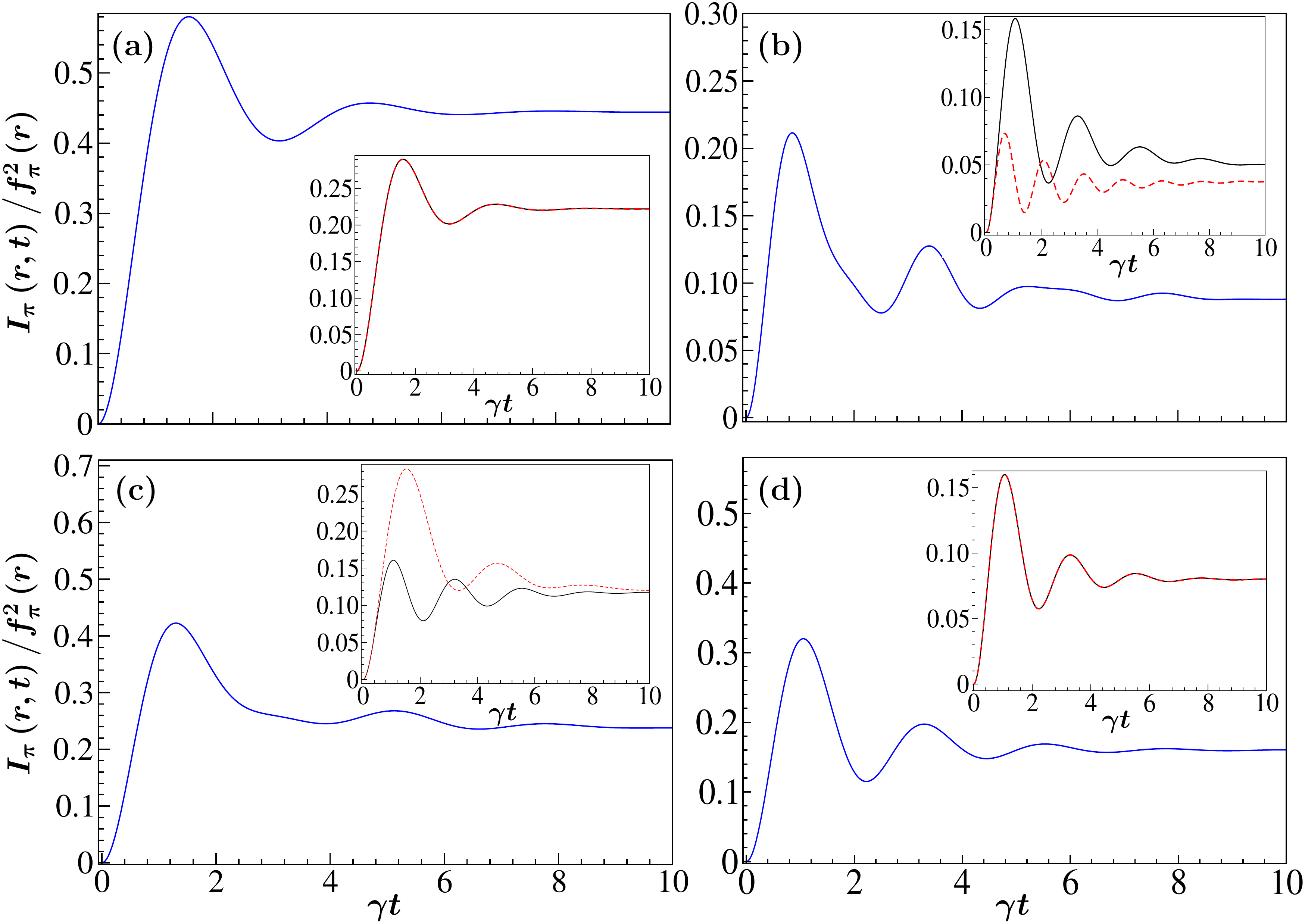} 
\caption{\label{fig:pop_intensityW} 
Fluorescence intensity (arb. units) of the $\pi$ transitions with equal initial 
ground state populations, 
$\langle A_{33} (0) \rangle = \langle A_{44} (0) \rangle = 1/2$ and 
$\langle A_{11} (0) \rangle = \langle A_{22} (0) \rangle = 0$.  
The other parameters are as in Fig.~2: $\Omega=\gamma$ and 
(a) $\Delta = \delta=0$; (b) $\Delta = 2 \gamma$, $\delta= -2\gamma$; 
(c) $\Delta = \delta= -2\gamma$; (d) $\Delta =-2 \gamma$, 
$\delta= -4\gamma$. The insets show the populations of the excited states: 
$\langle A_{11} (t) \rangle$ (solid-black), $\langle A_{22} (t) \rangle$ 
(dashed-red). } 
\end{figure} 

Just adding the excited state populations with the atom initially in the 
single state $|3\rangle$ in Eq.~(\ref{eq:time_Intensity_pi}) gives simply 
$I_{\pi} (\mathbf{r}, \hat{t}) = f_{\pi}^2(r) \langle A_{11} (\hat{t}) \rangle$, 
i.e., without the contribution of $\langle A_{22} (\hat{t}) \rangle$. More 
interesting is the case where the initial condition is 
$\langle A_{33} (0) \rangle = \langle A_{44} (0) \rangle = 1/2$, shown in 
Fig.~\ref{fig:pop_intensityW} (see the populations 
$\langle A_{11} (t) \rangle$ and $\langle A_{22} (t) \rangle$ in the insets). 
The modulation in the intensity is reminiscent of the quantum beats from 
spontaneous decay in multilevel systems \cite{FiSw04,FiSw05}, but with 
a nonzero steady state. Basically, the beats are due to the inability to tell 
if photons come from one $\pi$ transition or the other. The main 
requirement is that both ground states are initially nonzero, ideally equal 
\cite{FiSw04} (see Appendix \ref{beats_intensity}).  

\begin{figure}[t]
\includegraphics[width=8.5cm,height=8cm]{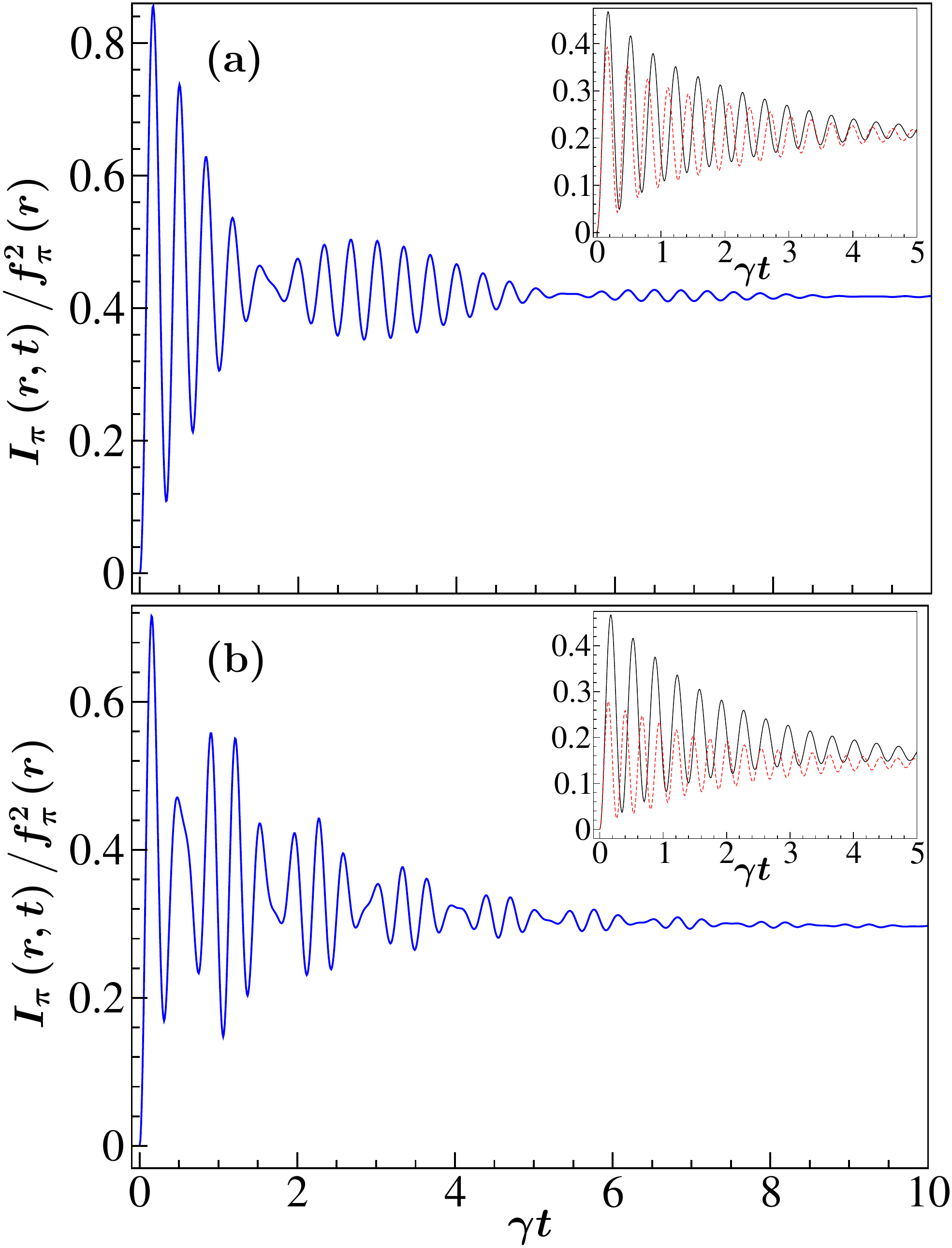} 
\caption{\label{fig:pop_intensity}  
Fluorescence intensity (arb. units) for $\Omega = 9\gamma$, $\Delta =0$, 
and (a) $\delta = -8\gamma$ and (b) $\delta = -15\gamma$. The insets 
show the excited state populations $\langle A_{11} \rangle$ (solid-black) and 
$\langle A_{22} \rangle$ (dotted-red). The initial conditions are  
$\langle A_{33} (0) \rangle =\langle A_{44} (0) \rangle = 1/2$, 
$\langle A_{11} (0) \rangle = \langle A_{22} (0) \rangle = 0$.   } 
\end{figure}

Still more interesting, though, is the case of strong resonant laser and 
magnetic fields, $\Omega,\delta \gg \gamma$, such that the laser gets 
detuned somewhat far from the $|2\rangle -|4\rangle$ resonance 
frequency, shown in Fig.~\ref{fig:pop_intensity}, with the populations 
$\langle A_{11} (t) \rangle$ and $\langle A_{22} (t) \rangle$ in the insets. 
Remarkably, beats with well-defined wave-packets are developed due to 
the interference of the fluorescence of both $\pi$ transitions with close 
Rabi frequencies, with clearly distinct average and modulation frequencies. 
With a larger difference Zeeman splitting, the pulsations become shorter 
and tend to scramble the signal. 

Save for the decay, these beats look more like those seen in introductory 
wave physics, described by a modulation \textit{and} an average frequency, 
unlike the beats from spontaneous emission or weak resonance 
fluorescence from two or more closely separated levels. Henceforth, we 
(mainly) reserve the moniker \textit{beats} to those due to strong applied 
fields. Further analyses of the beats are given in the following Sections, 
as they also show up in two-time correlations with particular features. 

As an aside, we note that for the $\sigma$ transitions we have 
\begin{subequations}
\begin{eqnarray} 
I_{\sigma}(\mathbf{r}, \hat{t}) &=&  \langle \hat{E}_{\sigma}^- (\mathbf{r}, \hat{t}) 
	\cdot  \hat{E}_{\sigma}^+  (\mathbf{r}, \hat{t}) \rangle \nonumber \\
	&=& f_{\sigma}^2(r) [\langle A_{23} (\hat{t}) A_{32} (\hat{t}) 
	+ A_{14} (\hat{t}) A_{41} (\hat{t}) \rangle  ]     \nonumber \\ 
	&=& f_{\sigma}^2(r) [\langle A_{11} (\hat{t}) + A_{22} (\hat{t}) \rangle ] 
	 \label{eq:time_intensity_sigma} , \\ 
I_{\sigma}^{st} &=& f_{\sigma}^2(r) \left[ \alpha_{11} + \alpha_{22} \right] , 
	\label{eq:steady_intensity_sigma} 
\end{eqnarray} 
\end{subequations}
that is, also showing beats but with intensity twice that of the $\pi$ 
transitions, given that $f_{\sigma}^2(r) = 2 f_{\pi}^2(r)$.

\subsection{Intensity Fluctuations} 
A different angle to interference in resonance fluorescence is found by 
considering fluctuations of the field even though the total intensity is 
not affected by interference. Here, we introduce the field's intensity 
in terms of atomic fluctuation operators 
$\Delta A_{jk} = A_{jk} - \langle A_{jk} \rangle_{st}$, such that  
\begin{eqnarray}
\langle A_{kl} A_{mn} \rangle &=& \alpha_{kl} \alpha_{mn} 
	+\langle \Delta A_{kl} \Delta A_{mn} \rangle  . \label{eq:splitop}
\end{eqnarray}
Only the $\pi$ transitions have nonzero coherence terms 
($\alpha_{13}, \alpha_{24} \neq 0$). Fluorescence in the $\sigma$ 
transitions is fully incoherent ($\alpha_{14} = \alpha_{23} =0$), its 
intensity given by Eq.~(\ref{eq:steady_intensity_sigma}) so, in the 
remainder of this Section, we deal only with the $\pi$ transition. 

From Eqs.~(\ref{eq:steady_intensity_pi}) and (\ref{eq:splitop}) we write the 
steady-state intensity in terms of products of dipole and dipole fluctuation 
operator expectation values, 
\begin{eqnarray}
I_{\pi}^{st}(\mathbf{r}) 
	&=& f_{\pi}^2(r) \left[ I_{\pi,0}^{coh} + I_{\pi,0}^{inc} 
	+ I_{\pi,cross}^{coh} + I_{\pi,cross}^{inc} \right] , 
\end{eqnarray}
where 
\begin{subequations}
\begin{eqnarray} 	
I_{\pi,0}^{coh} 
	&=& |\langle A_{13} \rangle_{st}|^2 +|\langle A_{24} \rangle_{st}|^2 , \\ 
I_{\pi,0}^{inc} &=& \langle \Delta A_{13} \Delta A_{31} \rangle 
    + \langle \Delta A_{24} \Delta A_{42} \rangle     , \\ 
I_{\pi,cross}^{coh} 
    &=& -\langle A_{13} \rangle_{st} \langle A_{42} \rangle_{st} 
    - \langle A_{24} \rangle_{st} \langle A_{31} \rangle_{st}    \nonumber  \\ 
    &=& -2 \mathrm{Re} \left( 
    \langle A_{13} \rangle_{st} \langle A_{42} \rangle_{st} \right) , \\ 
I_{\pi,cross}^{inc} &=& - \langle \Delta A_{13} \Delta A_{42} \rangle 
    - \langle \Delta A_{24} \Delta A_{31} \rangle    \nonumber  \\ 
    &=& -2 \mathrm{Re} \left( 
    \langle \Delta A_{13} \Delta A_{42} \rangle \right) .
\end{eqnarray}
\end{subequations}
Superindices $coh$ and $inc$ stand, respectively, for the coherent 
(depending on mean dipoles) and incoherent (depending on noise terms) 
parts of the emission. Subindex $0$ stands for terms with the addition of 
single transition products, giving the total intensity, while subindex $cross$ 
stands for terms with products of operators of the two $\pi$ transitions, the 
steady state interference part of the intensity. In terms of atomic expectation 
values, these intensities are: 
\begin{subequations} 	\label{eq:meanIntensities}
\begin{eqnarray}
I_{\pi,0}^{coh} &=& |\alpha_{13}|^2 +|\alpha_{24}|^2  \\ 
	&=& \frac{\Omega^2}{4D^2} \left[ \frac{\gamma^2}{2} 
	+\Delta^2 +(\delta-\Delta)^2 \right]   ,	\nonumber \\
I_{\pi,0}^{inc} 
	&=& \alpha_{11} +\alpha_{22} -|\alpha_{13}|^2 -|\alpha_{24}|^2     \\ 
	&=& \frac{\Omega^2}{D^2} \left[ 2\Omega^2 -\frac{\gamma^2}{4} 
	-\Delta^2 -\delta^2 \right]   ,	\nonumber \\
I_{\pi,cross}^{coh} 
 	&=& -2 \mathrm{Re} \left( \alpha_{13}\alpha_{42}  \right) 	 \\ 
	&=& \frac{\Omega^2}{2D^2} \left[ \frac{\gamma^2}{4} 
	+\Delta (\Delta-\delta) \right]   ,	\nonumber \\
I_{\pi,cross}^{inc} &=& 2 \mathrm{Re} \left( \alpha_{13}\alpha_{42}  \right) 
	= - I_{\pi,cross}^{coh} 	, 
\end{eqnarray}
\end{subequations}
The sum of these terms is, again, the total intensity, 
Eq.~(\ref{eq:time_Intensity_pi}). As usual in few-level resonance 
fluorescence, the coherent and incoherent intensities are similar only in 
the weak field regime, $\Omega \leq \gamma$; for strong fields, 
$\Omega \gg \gamma$, almost all the fluorescence is incoherent and    
here, in particular, $I_{\pi,0}^{inc} \gg I_{\pi,cross}^{inc}$, that is, the 
noninterference part becomes dominant.  

\subsection{Degree of Interference - Coherent Part} 
\begin{figure}[t]
\includegraphics[width=8.5cm,height=6cm]{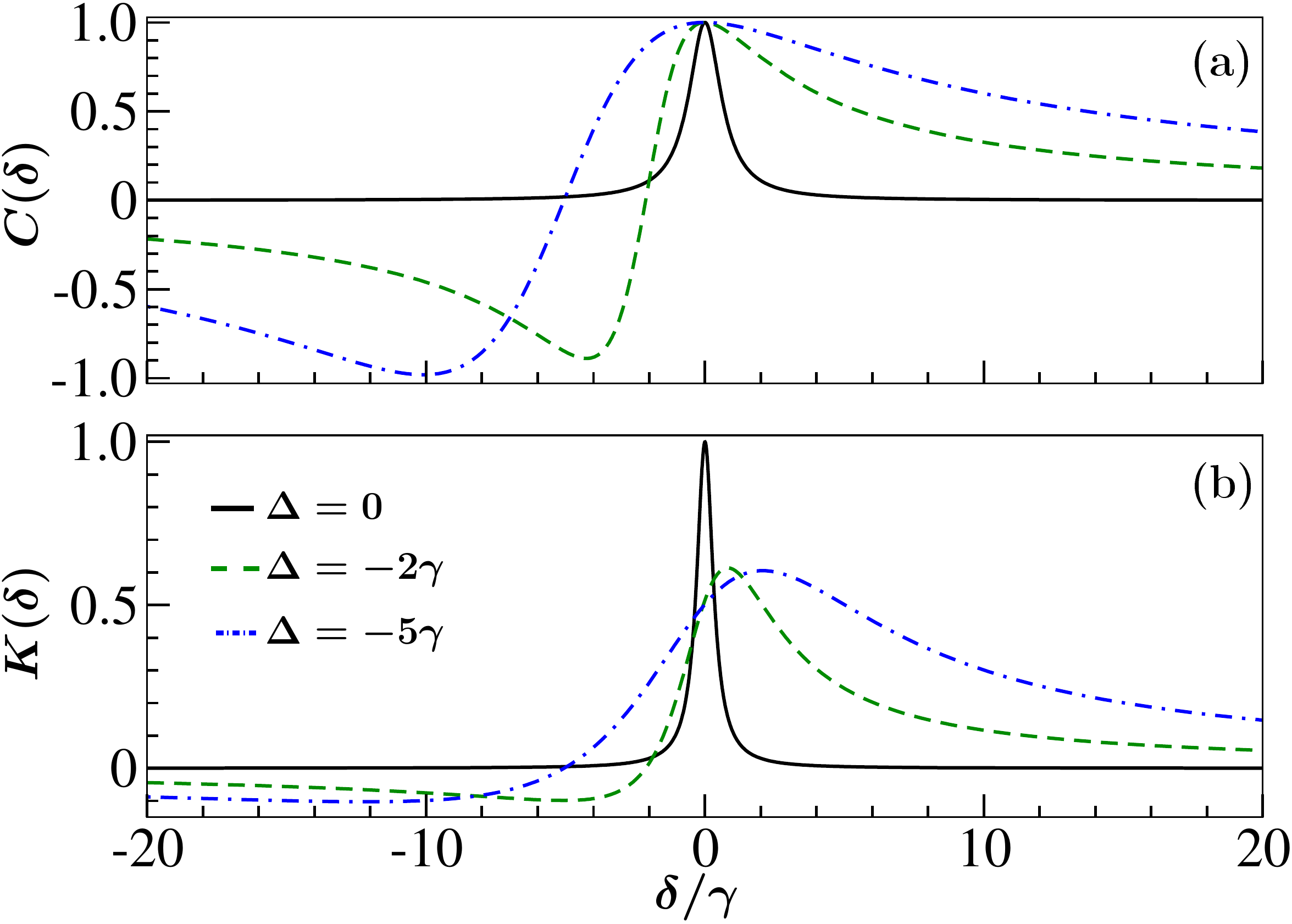} 
\caption{\label{fig:interferenceTerm} 
Relative weight of the interference terms $C(\delta)$ (a) and $K(\delta)$ (b). 
In (b) $\Omega=\gamma/4$. For $^{198} \mathrm{Hg}^+$, $\delta \leq 0$. }  
\end{figure}

In Ref.~\cite{KiEK06b}, a measure of the effect of interference in the 
coherent part of the intensity was defined as  
\begin{eqnarray} 	
I_{\pi,0}^{coh} +I_{\pi,cross}^{coh} &=& I_{\pi,0}^{coh} (1+C(\delta)) , 
	\nonumber \\ 
C(\delta) = \frac{ I_{\pi,cross}^{coh} }{ I_{\pi,0}^{coh} } 
	&=& \frac{ \gamma^2/4 +\Delta(\Delta -\delta) }
		{ \gamma^2/4 +\delta^2/4 +(\Delta -\delta/2)^2 } ,
		\label{eq:C-interf}
\end{eqnarray}
independent of the Rabi frequency, shown in Fig.~\ref{fig:interferenceTerm}(a). 

Some special cases are found analytically: 
\begin{subequations}
\begin{eqnarray}
C(0) &=&1 , \qquad \delta = 0 , \\ 
C(\delta_0) &=& 0 , \qquad \delta_0 = \Delta [1+ (\gamma/2\Delta)^2] , \\ 
C(\delta_{min}) &=& \frac{-1}{1+ \gamma^2/2\Delta^2} , 
	\qquad \delta_{min} = 2\Delta [1+ (\gamma/2\Delta)^2] , \nonumber \\ \\
C(\delta_{1/2}^{\pm}) &=& 1/2 , \qquad 
	\delta_{1/2}^{\pm} = -\Delta \pm \sqrt{ 3\Delta^2 +(\gamma^2/2)  } . 
	\nonumber \\
\end{eqnarray}
In the degenerate case, $C(\delta=0)=1$ means perfect constructive 
interference. That is because, at $\delta=0$, all transitions share the same 
reservoir environment. Increasing $\delta$ the reservoir overlap decreases, 
and so is the interference. Negative values of $C$ indicate destructive 
interference; its minimum is given by $\delta_{min}$. For large detunings, 
$\Delta^2 \gg \gamma^2$, we have 
\begin{eqnarray} 
\delta_0 = \Delta , 	\qquad  \delta_{min} = 2\Delta , 	 \qquad 
	 \delta_{1/2}^{\pm} = - \Delta \pm \sqrt{3} \,|\Delta|  . \nonumber \\
\end{eqnarray}
\end{subequations} 
We have used the special cases $\delta = \{0, \delta_0, \delta_{min} \}$ 
as a guide to obtain many of the figures in this paper. 

Although $C(\delta)$ is independent of Rabi frequency, it is important to 
recall that the coherent intensity is small in the strong laser field regime, 
Eqs.~(\ref{eq:meanIntensities}a,c), so it virtually has no role in the 
formation of beats. 

\subsection{Degree of Interference - Incoherent Part} 
Likewise, we define a measure, $K(\delta)$, of the effect of interference
 in the intensity's incoherent part,   
\begin{eqnarray} 	
I_{\pi,0}^{inc} +I_{\pi,cross}^{inc} &=& I_{\pi,0}^{inc} (1+K(\delta)) 	, 
	\nonumber \\ 
K(\delta) = \frac{ I_{\pi,cross}^{inc} }{ I_{\pi,0}^{inc} } 
	&=& \frac{ \gamma^2/4 +\Delta(\Delta -\delta) }
		{ 2\left[ \gamma^2/4 +\delta^2 +\Delta^2 -2\Omega^2 \right] } .
		\label{eq:K-interf}
\end{eqnarray}
Unlike $C(\delta)$, $K(\delta)$ also depends on the Rabi frequency as 
$\Omega^{-2}$, since fluctuations increase with laser intensity. Special 
cases are: 
\begin{subequations}
\begin{eqnarray}
K(0) &=& \frac{ \gamma^2/4 +\Delta^2 }
		{ 2\left[ \gamma^2/4 +\Delta^2 -2\Omega^2 \right] } , 
		\qquad \delta = 0 , \\ 
K(\delta) &=& 0 , \quad \delta = \Delta +\frac{\gamma^2}{4\Delta}  \quad 
	\mathrm{or} \quad \Omega \gg \gamma, \Delta, \delta . \qquad  
\end{eqnarray}
\end{subequations} 
The behavior of $K(\delta)$ with $\Delta$ is more subtle. It is basically 
required that $\Delta \sim \Omega$ in order to preserve the shape seen 
in Fig.~\ref{fig:interferenceTerm}(b), in which case the minima for 
$C(\delta)$ and  $K(\delta)$ are very similar. On-resonance, for example, 
$\Omega$ should be no larger than $0.35 \gamma$. Also, we can infer 
that the beats are little affected by the interference term unless 
$\Delta \gtrsim \Omega \gg \gamma$.  

\section{Two-Time Dipole Correlations and Power Spectrum} 
We have talked about the $\pi$ transitions evolving with different 
frequencies and the resulting quantum beats in the intensity with little 
quantitative analysis. The reason for delaying the analysis lies in that it is 
difficult to obtain general analytic solutions from the eight-equation system. 
Fortunately, in the strong field regime, the dressed system approach allows 
us to obtain very good approximate expressions, which then help us to 
discern the origin and positions of the peaks of the spectrum from the 
transitions among the dressed states \cite{KiEK06b}. 

The resonance fluorescence spectrum of the $J=1/2 \to J=1/2$ atomic 
system was first considered in \cite{PoSc76} and then very thoroughly in 
\cite{KiEK06a,KiEK06b}. In the strong field regime and large Zeeman 
splittings, a spectrum emerges with a central peak and not one 
\cite{Mollow69} but two pairs of sidebands. A major result of our paper is 
the observation that the closeness of the side peaks makes the components 
interfere, causing quantum beats with well-defined mean and modulation 
frequencies. 

The stationary Wiener-Khintchine power spectrum is given by the Fourier 
transform of the field autocorrelation function 
\begin{eqnarray}
S_{\pi}(\omega) &=& \mathrm{Re} \int_0^{\infty} d\tau e^{-i \omega \tau} 
	\langle \hat{E}_{\pi}^- (0) \hat{E}_{\pi}^+ (\tau) \rangle  ,
\end{eqnarray}
such that $\int_{-\infty}^{\infty} S_{\pi}(\omega) d\omega = I_{\pi}^{st}$. 
By writing the atomic operators in Eq.~(\ref{eq:field-atomOps}) as 
$A_{jk} (t) = \alpha_{jk} +\Delta A_{jk} (t)$, we separate the spectrum in 
two parts: a coherent one, 
\begin{eqnarray} 	\label{eq:ScohPi}
S_{\pi}^{coh}(\omega) 
&=&  \mathrm{Re} \int_0^{\infty}  e^{-i \omega \tau} d\tau 
	\left[ I_{\pi,0}^{coh} + I_{\pi,cross}^{coh} \right]	\nonumber \\
&=& \pi \left[ I_{\pi,0}^{coh} + I_{\pi,cross}^{coh} \right]  \delta(\omega) 
	\nonumber \\ 
&=& \frac{\pi \Omega^2}{D^2} \left[ \frac{\gamma^2}{4} 
	+\left( \Delta -\frac{\delta}{2} \right)^2 \right] \delta(\omega),
\end{eqnarray}
due to elastic scattering, where $I_{\pi,0}^{coh}$ and $I_{\pi,cross}^{coh}$ 
are given by Eqs.~(\ref{eq:meanIntensities}) (a) and (c), respectively; and  
an incoherent part, 
\begin{eqnarray} 	\label{eq:Sinc}
S_{\pi}^{inc}(\omega) &=& \mathrm{Re} \int_0^{\infty} d\tau e^{-i \omega \tau} 
  \langle \Delta \hat{ E}_{\pi}^- (0) \Delta \hat{E}_{\pi}^+ (\tau) \rangle  , 
  \nonumber 
\end{eqnarray}
specifically, 
\begin{eqnarray} 	\label{eq:SincPi}
S_{\pi}^{inc}(\omega) &=&  \mathrm{Re} \int_0^{\infty} d\tau e^{-i \omega \tau} 
	\left[ \langle \Delta A_{13}(0) \Delta A_{31}(\tau) \rangle 
	\right. \nonumber \\
&&	+\langle \Delta A_{24}(0) \Delta A_{42}(\tau) \rangle  
	-\langle \Delta A_{13}(0) \Delta A_{42}(\tau) \rangle  	\nonumber \\
&& \left.	-\langle \Delta A_{24}(0) \Delta A_{31}(\tau) \rangle  \right] , 
\end{eqnarray}
due to atomic fluctuations. An outline of the numerical calculation is given 
in  Appendix \ref{ap:matrixSol}.  

\begin{figure}[t]
\includegraphics[width=8.5cm,height=5cm]{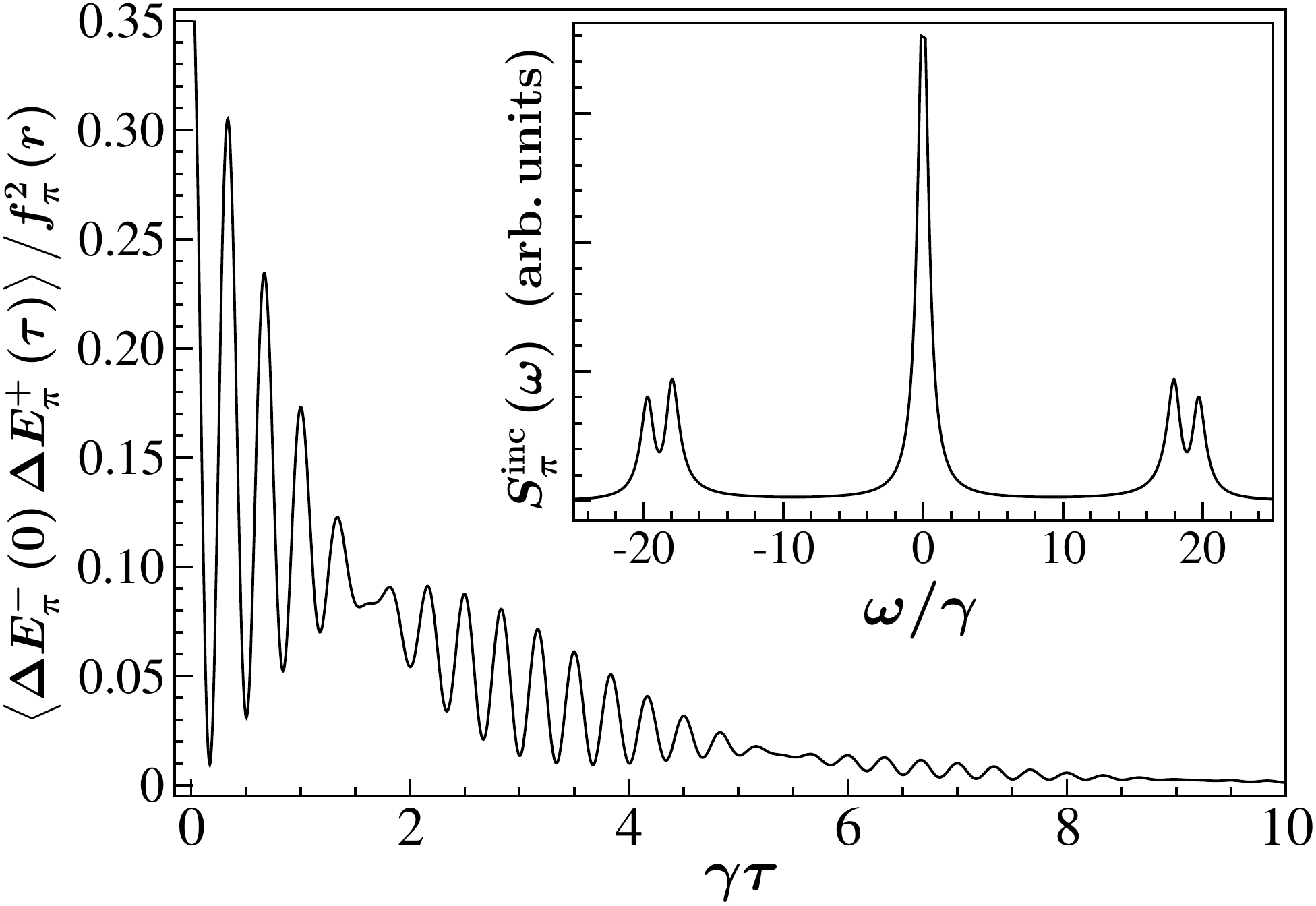} 
\caption{\label{spectrum} 
Dipole correlation function 
$\langle \Delta \hat{E}_{\pi}^- (0) \Delta \hat{E}_{\pi}^+ (\tau) \rangle$ for 
$\Omega = 9\gamma$, $\delta = -8\gamma$, and $\Delta =0$. The inset 
shows the corresponding incoherent spectrum $S_{\pi}^{inc}(\omega)$.   }
\end{figure}
The dipole correlation 
$\langle \hat{E}_{\pi}^- (0) \hat{E}_{\pi}^+ (\tau) \rangle$ and the 
corresponding incoherent spectrum in the strong driving regime and 
strong nondegeneracy (large $\delta$) are shown in Fig.~\ref{spectrum}. 
The sidebands are localized at the generalized Rabi frequencies
 $\pm \Omega_1$ and $\pm \Omega_2$, given by
\begin{subequations}
\begin{eqnarray} 	\label{eq:Rabis} 
\Omega_1 &=& \mathcal{E}_1^+ -\mathcal{E}_1^- 
	=  \sqrt{ 4\Omega^2 +\Delta^2 }  , \\
\Omega_2 &=& \mathcal{E}_2^+ -\mathcal{E}_2^- 
	=  \sqrt{ 4\Omega^2 +(\delta -\Delta)^2 }  , 
\end{eqnarray}
\end{subequations}
where 
\begin{subequations}
\begin{eqnarray}  	\label{eq:H_eigenvalues}
\mathcal{E}_1^{\pm} 
	&=& - \frac{\Delta}{2}  \pm \frac{1}{2} \sqrt{ 4\Omega^2 +\Delta^2 } , \\ 
\mathcal{E}_2^{\pm} 
	&=& B_{\ell} + \frac{ \delta -\Delta}{2}  
		\pm \frac{1}{2} \sqrt{ 4\Omega^2 +(\delta -\Delta)^2 }    , 
\end{eqnarray}
\end{subequations}
are eigenvalues of the Hamiltonian (\ref{eq:Hamiltonian}). Due to the 
spontaneous decays, these frequencies would have to be corrected, but 
they are very good in the relevant strong field regime. Indeed, we notice that 
$\Omega_1$ and $\Omega_2$ are very close to the imaginary parts of the 
numerical eigenvalues $\lambda_{2,3}$ and $\lambda_{4,5}$, respectively, 
of matrix $\mathbf{M}$, as shown in Table~I. 

\begin{table}[t] 		
\caption{ \label{tb:NEigenvalues} Eigenvalues of matrix 
$\mathbf{M}/\gamma$ and initial conditions of the correlations in 
Eq.~(\ref{eq:SincPi}) for $\Omega =9\gamma$ and $\Delta =0$. }
\begin{ruledtabular}
\begin{tabular}{ccc} 		
\textrm{Eigenvalues}  & $\delta = -8\gamma$	
	& $\delta =-15\gamma$ 			\\ \hline
$\lambda_1$  & $-0.749386 +0i$ & $-0.836531 +0i$ 	 		\\ 
$\lambda_2$ &  $-0.583099 -18.0094 i$ & $-0.583308 -17.9981 i $	\\ 	
$\lambda_3$ & $-0.583099 +18.0094 i$ 	& $-0.583308 +17.9981 i$       \\ 	
$\lambda_4$ &  $-0.569785 -19.6808 i$ & $-0.5492 -23.4257 i$  	 \\ 	
$\lambda_5$ &   $-0.569785 +19.6808 i$ & $-0.5492 +23.4257 i$ 	 \\ 	
$\lambda_6$  & $-0.5 +0i$ & $-0.5 +0i$ 	 	\\ 	
$\lambda_7$ & $-0.444846 +0i$ & $-0.398452 +0i$ 	\\  
$\lambda_8$ & $0 +0i$ & $0 +0i$ 	\\ 	\hline \hline  
\textrm{Initial condition} & & \\ \hline 
$\langle \Delta A_{13} \Delta A_{31} \rangle$ & $0.20836 +0i$ 
	& $0.14734 +0i$ \\ 
$\langle \Delta A_{24} \Delta A_{42} \rangle$ & $0.174014 +0i$ 
	& $0.086982 +0i$ \\ 
$\langle \Delta A_{13} \Delta A_{42} \rangle$ & $0.000134 + 0.002146i$ 
	& $0.000067 +0.002011i$ \\ 
$\langle \Delta A_{24} \Delta A_{31} \rangle$ & $0.000134 - 0.002146i$ 
	& $0.000067 -0.002011i$  
\end{tabular}
\end{ruledtabular}
\end{table}

The beats are the result of the superposition of waves at the frequencies 
$\Omega_1$ and $\Omega_2$ of the spectral sidebands, with average 
frequency
\begin{eqnarray} 	\label{eq:beatfreq}
\Omega_{av} = \frac{\Omega_2 +\Omega_1}{2} 
	&=& \frac{ \sqrt{ 4\Omega^2 +(\delta -\Delta)^2 }
	+ \sqrt{ 4\Omega^2 +\Delta^2 }	}{2} , \nonumber \\
\end{eqnarray}
and beat or modulation frequency
\begin{eqnarray} 	\label{eq:beatwp}
\Omega_{beat} = \frac{\Omega_2 -\Omega_1}{2} 
	&=& \frac{ \sqrt{ 4\Omega^2 +(\delta -\Delta)^2 }
	- \sqrt{ 4\Omega^2 +\Delta^2 }	}{2} . \nonumber \\
\end{eqnarray}

Now, we  can identify the origin of the beats in the time-dependent intensity, 
Eq.~(\ref{eq:time_Intensity_pi}), since the excited state populations 
$\langle A_{11} (t) \rangle$ and $\langle A_{22} (t) \rangle$ oscillate at the 
generalized Rabi frequencies $\Omega_1$ and $\Omega_2$, respectively, 
with initial conditions given by a nonzero superposition of the ground state 
populations at $t=0$. In the case of the dipole correlation 
$\langle \hat{E}_{\pi}^- (0) \hat{E}_{\pi}^+ (\tau) \rangle$, however, the 
initial conditions are given in terms of products of stationary atomic 
expectation values (most of them the coherences 
$\alpha_{13},\alpha_{24}$), which become very small for 
$\Omega \gg \gamma$. Thus, as seen in the bottom part of Table~I, the 
terms $\langle \Delta A_{13}(0) \Delta A_{31}(\tau) \rangle$ and 
$\langle \Delta A_{24}(0) \Delta A_{42}(\tau) \rangle$ are much larger than 
the cross terms $\langle \Delta A_{13}(0) \Delta A_{42}(\tau) \rangle$ and 
$\langle \Delta A_{24}(0) \Delta A_{31}(\tau) \rangle$, so the beats are 
basically due to the interference of those dominant terms. The strong 
damping of the beats, compared to those of the intensity, is due to the real 
eigenvalues $\lambda_1,\lambda_6,\lambda_7$. The first eigenvalue, 
$\lambda_1 \sim - 3\gamma/4$, in particular, is like that of the 2LA and other 
cases in RF. Also, if $\Delta \neq 0$ more eigenvalues would be complex, 
but this is an unnecessary complication in understanding the emergence of 
beats.

We also note that when $\delta = 2\Delta$, $\Omega_1$ equals 
$\Omega_2$, thus the two sidebands merge into a single one, thus the 
beats disappear, also explaining why the populations 
$\langle A_{11}(\tau) \rangle$ and $\langle A_{22}(\tau) \rangle$
are out of phase, as in the degenerate case.

\section{Photon-Photon Correlations}  
The standard method to investigate intensity fluctuations of a light source 
uses Brown-Twiss photon-photon correlations \cite{HBT,Glauber-g2}. The 
conditional character of this type of measurement makes it nearly free of 
detector inefficiencies, unlike single-detector measurements of  
fluorescence. This makes it one of the most used techniques in quantum 
optics. In Ref.~\cite{DasAg08}, the effect of interference in the correlations 
of two photons from the $\pi$ transitions was studied, albeit only for the 
degenerate case and not too strong laser, finding a stronger damping of 
the Rabi oscillations than without interference. In this paper, we extend the 
photon correlations to the case of nondegenerate states. These correlations 
are defined as  
\begin{eqnarray}
g_{\pi}^{(2)} (\tau) = \frac{G_{\pi}^{(2)} (\tau)}{G_{\pi}^{(2)} (\tau \to \infty)}  
\end{eqnarray}
where, using Eq.~(\ref{eq:field-atomOps}) for the field operators, 
\begin{subequations}
\begin{eqnarray}
G_{\pi}^{(2)} (\tau) &=& \langle \hat{E}_{\pi}^- (0) \hat{E}_{\pi}^- (\tau) 
	 \hat{E}_{\pi}^+ (\tau) \hat{E}_{\pi}^+ (0)  \rangle  	\nonumber \\ 
 &=& f_{\pi}^4 (r) \langle  [ A_{13} (0) -A_{24} (0)] 	
 	[A_{11} (\tau) +A_{22} (\tau)  ]  		\nonumber \\ 
	&& \times [A_{31} (0) -A_{42} (0)]  \rangle , 	\label{eq:G2full}
\end{eqnarray}
and 
\begin{eqnarray}
G_{\pi}^{(2)} (\tau \to \infty) &=& \left( I_{\pi}^{st}  \right)^2
	= f_{\pi}^4 (r) \left( \alpha_{11} + \alpha_{22} \right)^2  
\end{eqnarray}
is the normalization factor. Note that the full photon correlation, 
Eq.(\ref{eq:G2full}), is not just the sum of the correlations of the separate  
$\pi$ transition, but it contains cross terms mixing both paths, adding to the 
total interference, unlike the total intensity.

The terms of the correlation obey the quantum regression formula, 
Eqs.(\ref{eq:EqQRF}) and (\ref{eq:SolQRF}). Then, from Eq.(\ref{eq:AjkRe}), 
$G_{\pi}^{(2)} (\tau)$ is further reduced as 
$\langle A_{13}(0)  A_{jk} (\tau) A_{42} (0) \rangle 
= \langle A_{24}(0) A_{jk} (\tau) A_{31} (0) \rangle =0$. 
We are then left with four terms, 
\begin{eqnarray}
G_{\pi}^{(2)} (\tau) &=& f_{\pi}^4 (r) \left\{ \langle A_{13} (0) 
	[A_{11} (\tau) +A_{22} (\tau)]  A_{31} (0)   \rangle   \right. \nonumber \\ 
	&& \left. +\langle A_{24} (0) [A_{11} (\tau) +A_{22} (\tau)] 
	A_{42} (0) \rangle \right\} .
\end{eqnarray} 
\end{subequations}
As usual in single-atom resonance fluorescence, the correlation shows the 
ubiquitous, nonclassical, antibunching effect, $g_{\pi}^{(2)} (0) =0$, that is, 
a single atom cannot emit two photons simultaneously. 
\begin{figure}[t]
\includegraphics[width=8.5cm,height=6cm]{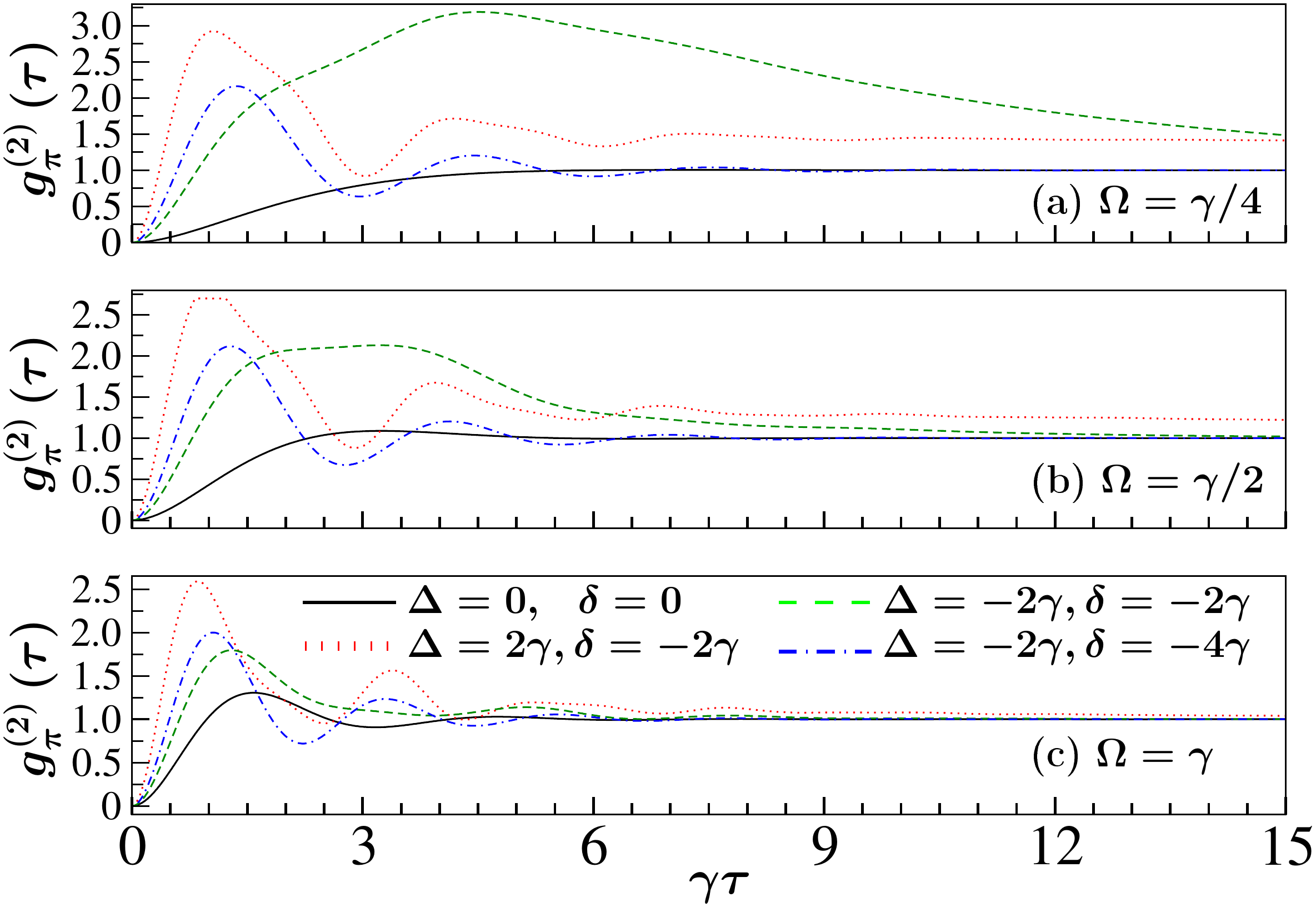} 
\caption{\label{fig:g2} 
Photon correlations for (a) $\Omega = \gamma/4$, (b) 
$\Omega = \gamma/2$, and (c) $\Omega = \gamma$. The pairs of 
detunings ($\Delta,\delta$) are the same as those in Fig.~\ref{fig:pops}.  }
\end{figure}
\begin{figure}[t]
\includegraphics[width=8.5cm,height=5cm]{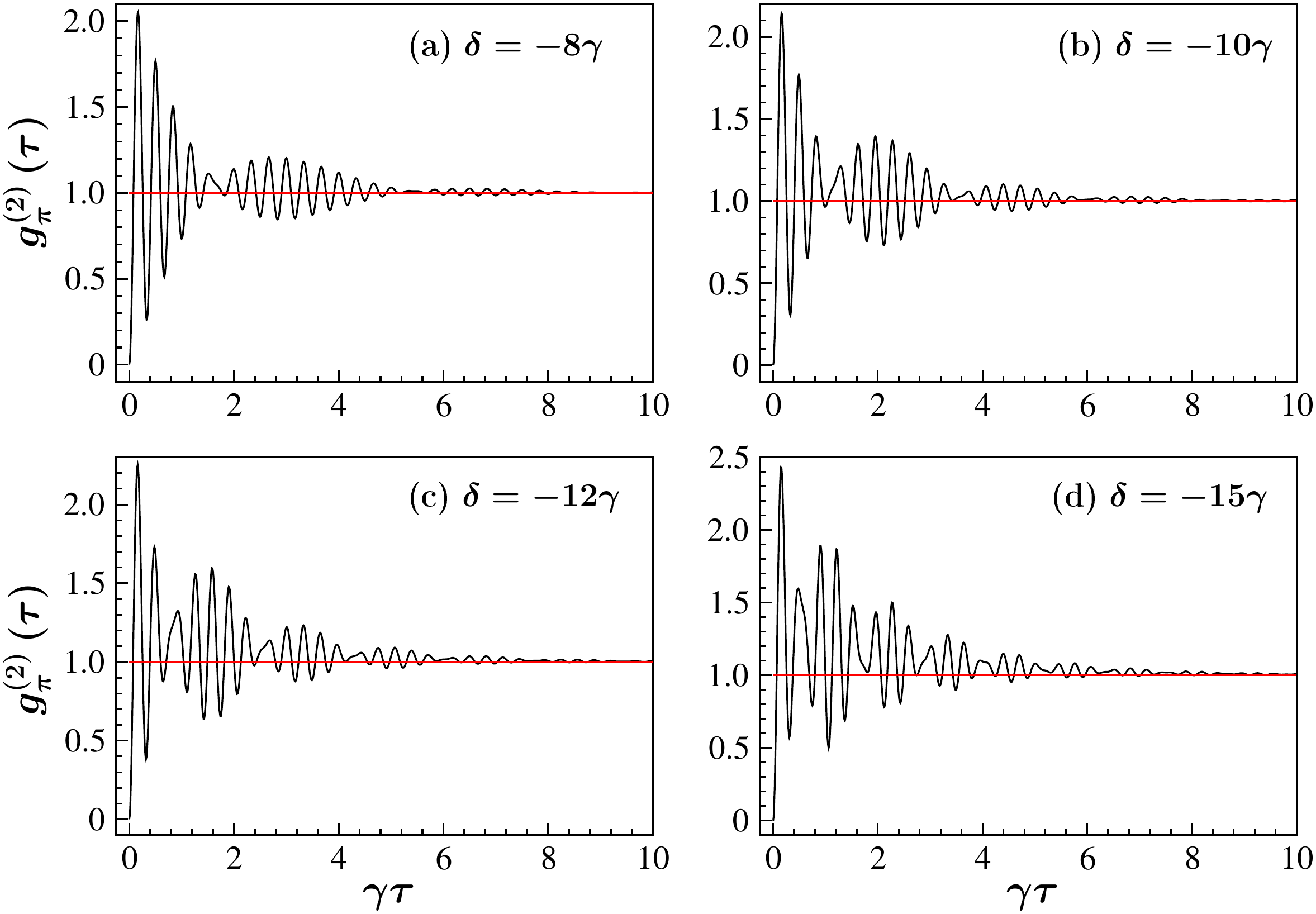} 
\caption{\label{fig:g2beats} 
Photon correlations in the strong field limit, $\Omega = 9\gamma$, 
$\Delta =0$, and large Zeeman splittings. The horizontal line helps to see 
that the wave packets are slightly above the long-time value  
$g_{\pi}^{(2)} (\infty) =1$. }   
\end{figure}

Figure~\ref{fig:g2} shows $g_{\pi}^{(2)} (\tau)$ for moderately strong drivings, 
$\Omega =\gamma/4,\gamma/2,\gamma$, and the same sets of detunings 
$(\Delta,\delta)$ of Fig.~\ref{fig:pops}. In the nondegenerate case, the multiple 
contributions and parameters cause a quite involved evolution. Notably, as 
seen in Fig.~\ref{fig:pops}, with nonzero laser detunings, one or the other 
ground state traps some of the population. In $g_{\pi}^{(2)} (\tau)$ this is 
manifested as a slow decay and lack of oscillations (see dashed green line). 
This trapping effect diminishes for increasing Rabi frequency, which would 
cause populations to evolve more evenly among states. We will see in the 
next Section that the trapping effect is related to a very narrow peak in the 
spectra of quadratures.   
 
Now, Fig.~\ref{fig:g2beats} shows the case of strong driving and large 
difference Zeeman splitting, $\Omega,\delta \gg \gamma$,  Here, to 
simplify matters, we chose the  atom-laser detuning $\Delta=0$. As it 
happens with the time-dependent populations of the excited states 
(Fig.~\ref{fig:pops}), the dominant nonvanishing terms of $g^{(2)}(\tau)$, 
$\langle A_{13} (0) A_{11} (\tau) A_{31} (0) \rangle$ and 
$\langle A_{24} (0) A_{22} (\tau) A_{42} (0) \rangle$, evolve with close  
eigenvalues $\lambda_2,\lambda_3,\lambda_4,\lambda_5$, causing 
quantum beats. Note the lack of strong damping from the real eigenvalues 
$\lambda_1,\lambda_6,\lambda_7$. Also, with initial conditions depending 
on the stationary populations, there is no need to prepare an initial 
superposition atomic state, as for beats in the intensity. 

There are several effects resulting from the increase in the nondegeneracy 
factor $\delta$: (i) the larger number of visible wave packets; (ii) both 
average and beat frequencies approach one another, so the wave packets 
get shorter, containing very few of the fast oscillations, as seen in 
Fig.~\ref{fig:g2beats}(d); and (iii) the wavepackets are initially slightly lifted 
above the $g^{(2)}(\tau \to \infty) =1$ value. 

In the next Section, we investigate amplitude-intensity correlations, which 
share the main dynamical features observed in $g^{(2)}(\tau)$, with due 
differences in character, as the former are also analyzed in the frequency 
domain.

\section{Quadrature Fluctuations} 	
Quantum optics deals with the properties of the electromagnetic field. 
Intensity is the particle aspect of its nature, where its fluctuations are 
characterized by photon statistics and photon correlations. Amplitude and 
phase comprise the wave aspect, of which the squeezing effect is its most 
salient property. Squeezing is the reduction of noise in one quadrature 
below that of a coherent state at the expense of the other \cite{WaMi08}. 
It is usually measured by balanced homodyne detection (BHD), but low 
quantum detector efficiency degrades the weak squeezing produced in 
resonance fluorescence and cavity QED systems. One alternative our group 
has used is conditional homodyne detection (CHD) \cite{CCFO00,FOCC00}, 
which correlates a quadrature amplitude on the cue of an intensity 
measurement. CHD measures a third-order amplitude-intensity correlation 
(AIC) which, in the weak driving limit, is reduced to the second-order one,  
which allows for measuring squeezing, nearly free of detector inefficiencies. 

While the original goal of CHD was to measure the weak squeezing in 
cavity QED \cite{CCFO00,FOCC00}, it was soon realized that nonzero 
third-order fluctuations of the amplitude provide clear evidence of 
non-Gaussian fluctuations and higher-order field nonclassicality. In the 
present work, the fluctuations are mainly of third-order due to excitation near 
and above saturation, and violate classical bounds. We thus explore 
the phase-dependent fluctuations under conditions of quantum interference, 
following our recent work on CHD in resonance fluorescence 
\cite{CaRG16,GCRH17,SaCa21}.  

\subsection{Quadrature Operators} 
To study the quadrature fluctuations, we define the field quadrature operator, 
\begin{eqnarray} 	\label{eq:fieldQuadOperator}
\hat{E}_{\pi,\phi} (\mathbf{r}, \hat{t}) 
	&=& \frac{1}{2} \left( E_{\pi}^- (\mathbf{r}, \hat{t}) e^{-i \phi} 
 	+ E_{\pi} ^+ (\mathbf{r}, \hat{t}) e^{i \phi} \right) 		\nonumber \\ 
	&=& f_{\pi}(r)   [S_{1, \phi}(\hat{t}) - S_{2, \phi}(\hat{t}) ] ,
\end{eqnarray}
where $\phi$ is the phase of the local oscillator required in homodyne-type 
measurements, and
\begin{subequations}
\begin{eqnarray}
S_{1, \phi} &=& \frac{1}{2} \left( A_{13} e^{-i \phi} + A_{31} e^{i \phi} \right) ,	\\  
S_{2, \phi} &=& \frac{1}{2} \left( A_{24} e^{-i \phi} + A_{42} e^{i \phi} \right) 	.
\end{eqnarray}
\end{subequations}

The mean quadrature field is given by 
\begin{eqnarray}
\langle \hat{E}_{\pi,\phi} \rangle_{st}  
	&=& \frac{f_{\pi}(r)}{2} \left[ \left( \alpha_{13} -\alpha_{24} \right) e^{-i \phi} 
	+\left( \alpha_{31} -\alpha_{42}  \right)e^{i \phi}  \right] 	\nonumber \\ 
&=& f_{\pi}(r) \mathrm{Re} 
	\left[ \left( \alpha_{13} -\alpha_{24} \right) e^{-i \phi} \right] 		\\ 
&=&  f_{\pi}(r) \mathrm{Re} \left[ 
\frac{ \Omega \left(\Delta +(i\gamma -\delta)/2 \right)}{D}  
	  e^{-i \phi} \right]  .		\nonumber 
\end{eqnarray}

Next we rewrite $\langle \hat{E}_{\pi,\phi} \rangle_{st}$ in terms of a mean 
quadrature and atomic fluctuation operators 
$\Delta A_{jk} = A_{jk} - \langle A_{jk} \rangle_{st}$,  
\begin{subequations}
\begin{equation}
\hat{E}_{\pi,\phi} (\mathbf{r}, \hat{t}) 
	= f_{\pi}(r) [\alpha_{\pi,\phi} +\Delta S_{\pi,\phi} (\hat{t})] , 
\end{equation}
where
\begin{eqnarray}
\alpha_{\pi,\phi}	&=& \frac{1}{2} ( \alpha_{31}- \alpha_{42}) e^{i \phi} 
	+ \frac{1}{2} ( \alpha_{13}- \alpha_{24}) e^{-i \phi} , \\ 
&=& \mathrm{Re} \left[ 
\frac{ \Omega \left(\Delta +(i\gamma -\delta)/2 \right)}{D}  
	  e^{-i \phi} \right]  ,		\nonumber 	\\
\Delta S_{\pi,\phi}&=& \frac{1}{2} ( \Delta A_{31}- \Delta A_{42}) e^{i \phi} 
	+ \frac{1}{2} ( \Delta A_{13}- \Delta_{24}) e^{-i \phi} .  \nonumber \\
\end{eqnarray}
\end{subequations}

A similar analysis for the $\sigma$ transitions tells us that this fluorescence 
is fully incoherent; its mean quadrature field vanishes because 
$\alpha_{14} =\alpha_{23} =0$. Higher-order fluctuations, like those from 
the amplitude-intensity correlations are small, if any.
  
\subsection{Amplitude-Intensity Correlations} 	
In CHD, a quadrature's field $E_{\phi}$ is measured by BHD on the cue 
of photon countings in a separate detector. This is characterized by a 
correlation between the amplitude and the intensity of the field, 
\begin{eqnarray} 	\label{eq:chd}
h_{\pi,\phi} (\tau) = \frac{H_{\pi,\phi} (\tau)}{H_{\pi,\phi} (\tau \to \infty)} , 
\end{eqnarray}
where 
\begin{subequations} 
\begin{eqnarray}
H_{\pi,\phi} (\tau) &=& \langle : 
	\hat{E}_{\pi}^- (0) \hat{E}_{\pi}^+ (0) \hat{E}_{\pi,\phi} (\tau) : \rangle , 
\end{eqnarray}
the dots $::$ indicating time and normal operator orderings, and 
\begin{eqnarray}
H_{\pi,\phi} (\tau \to \infty) &=& I_{\pi}^{st} \langle E_{\pi,\phi}  \rangle_{st} \\
	&=& f_{\pi}^3 (r) \left[ \alpha_{11} + \alpha_{22} \right] 
	\mathrm{Re} \left[ \left( \alpha_{13} -\alpha_{24} \right) e^{-i \phi} \right] 
	 \nonumber \\ 
&=&  f_{\pi}^3 (r) \frac{\Omega^3}{D^2} \mathrm{Re} \left[ 
	 \left(\Delta +(i\gamma -\delta)/2 \right)   e^{-i \phi} \right]  	\nonumber
\end{eqnarray}
\end{subequations}
is the normalization factor. 

For the sake of concreteness, in this Section, we limit our discussion to 
the out-of-phase quadrature, $\phi= \pi/2$, which is the one that features 
squeezing when $\Delta=0$. We do consider, however, squeezing in the 
in-phase quadrature $\phi=0$ in Sect. \ref{sec:variance} on the variance. 

In several atom-laser systems $h_{\pi,\phi} (\tau)$ has been shown to be 
time-asymmetric \cite{GCRH17,SaCa21}. This is not the case with the 
$J=1/2 \to J=1/2$ system, so we limit the analysis to positive intervals 
$\tau \geq 0$. Omitting the geometrical factor $f_{\pi}^3(r)$, which is later 
cancelled by normalization, we have 
\begin{eqnarray}
H_{\pi,\phi} (\tau) &=& \langle  \hat{E}_{\pi}^- (0) \hat{E}_{\pi,\phi} (\tau)  
	\hat{E}_{\pi}^+ (0)  \rangle 	\nonumber \\ 
&=& \mathrm{Re} \left\{ e^{-i\phi}  \langle  
	A_{13} (0) [A_{13} (\tau) -A_{24} (\tau) ] A_{31}(0) \right. \nonumber \\
	&& \left. +A_{24} (0) [A_{13} (\tau) -A_{24} (\tau)] A_{42}(0) 
		\rangle \right\} 	.  		\label{eq:cdh_atomic}
\end{eqnarray}
Note that $H_{\pi,\phi} (0) =0$, meaning that, like antibunching in 
$g^{(2)}(0)$, the atom has to build a new photon wavepacket after one  
has been emitted. 

The AIC suggests nontrivial behavior when we take dipole fluctuations 
into account, that is, when the atomic operators are split into their mean 
plus noise, $A_{jk} = \alpha_{jk} + \Delta A_{jk}$; upon substitution in Eq.~(\ref{eq:cdh_atomic}) we get 
\begin{eqnarray} 	\label{eq:chdsplit}
H_{\pi,\phi} (\tau) &=& I_{\pi}^{st} \langle E_{\pi,\phi}  \rangle_{st} 
	+ H_{\pi,\phi}^{(2)} (\tau) +H_{\pi,\phi}^{(3)} (\tau) , 	
\end{eqnarray}
or in normalized form as 
\begin{eqnarray} 	
h_{\pi,\phi} (\tau) &=& 
1+ \frac{H_{\pi,\phi}^{(2)} (\tau)}{I_{\pi}^{st} \langle E_{\pi,\phi}  \rangle_{st}}   
+\frac{H_{\pi,\phi}^{(3)} (\tau)}{I_{\pi}^{st} \langle E_{\pi,\phi}  \rangle_{st}} , 	
\end{eqnarray}
where
\begin{widetext}
\begin{eqnarray}
H_{\pi,\phi}^{(2)} (\tau) 
	&=&  2\mathrm{Re} \left[ \langle \hat{E}_{\pi}^+ \rangle_{st} 
	\langle \Delta \hat{E}_{\pi}^-(0) \Delta \hat{E}_{\pi,\phi}(\tau) 
		\rangle \right] 	  	\nonumber \\ 
&=& \mathrm{Re} \left\{ (\alpha_{31}- \alpha_{42}) \left[ 
	  \langle (\Delta A_{13} (0)  - \Delta A_{24} (0) \right) 
	\left( \Delta A_{13}(\tau)  - \Delta A_{24}(\tau) ) \rangle 
	 e^{-i\phi} 		\right.  \right. 	\nonumber \\ 
   && \left. \left. +  \langle (\Delta A_{13} (0)  - \Delta A_{24} (0) \right) 
	\left( \Delta A_{31}(\tau) \rangle - \Delta A_{42}(\tau) ) \rangle 
	 e^{i\phi} 		\right]  \right\}  ,
\end{eqnarray}
\begin{eqnarray}
H_{\pi,\phi}^{(3)} (\tau) &=& \langle \Delta \hat{E}_{\pi}^-(0) 
	\Delta \hat{E}_{\pi,\phi}(\tau) \Delta \hat{E}_{\pi}^+(0) \rangle   
	\nonumber \\ 
&=& 	\mathrm{Re} \left\{ e^{i\phi}   
	\langle \left[ \Delta A_{13} (0) - \Delta A_{24} (0) \right] 	
	 \left[ \Delta A_{31} (\tau) - \Delta A_{42} (\tau) \right] 	
	  \left[ \Delta A_{31} (0) - \Delta A_{42} (0) \right] \rangle  \right\} .
\end{eqnarray}
\end{widetext}
The initial conditions of the correlations are given in Appendix 
\ref{ap:matrixSol}. 

From $h_{\pi,\pi/2} (0) =0$ we can obtain analytically the initial values 
of the second- and third-order terms, 
\begin{eqnarray}
h_{\pi,\pi/2}^{(2)} (0) &=& 1- \frac{(2\Delta -\delta)^2 +\gamma^2}{2D} , \\
h_{\pi,\pi/2}^{(3)} (0) &=& \frac{(2\Delta -\delta)^2 +\gamma^2}{2D} -2 , 
\end{eqnarray}
where $D$ is given by Eq.~(\ref{eq:denominator}). 

Being the AIC a function of odd-order in the field amplitude we rightly 
expect a richer landscape than that of the intensity correlations, more so 
when one considers quantum interference. For instance, the correlation 
can take on not only negative values, breaking the classical bounds 
\cite{CCFO00,FOCC00}: 
\begin{subequations} 	\label{eq:ineq}
\begin{eqnarray} 	
0 &\leq& h_{\phi} (\tau) - 1 \leq 1 	\,, 	\\
| h_{\phi}^{(2)} (\tau) - 1 | &\leq& | h_{\phi}^{(2)} (0) - 1 | \leq 1 \,,   
	\label{eq:ineqb}
\end{eqnarray}
\end{subequations}
where the second line is valid only for weak fields such that 
$h_{\phi}^{(3)} (\tau) \sim 0$. These classical bounds are stronger criteria 
for the nonclassicality of the emitted field than squeezed light measurements, 
the more familiar probing of phase-dependent fluctuations. A detailed 
hierarchy of nonclassicality measures for higher-order correlation 
functions is presented in Refs.~\cite{ScVo05,ScVo06}. In 
Ref.~\cite{GCRH17} an inequality was obtained that considers the full 
$h_{\phi}(\tau)$ by calculating the AIC for a field in a coherent state, 
\begin{align} 	\label{eq:ineq_cs}
-1 \le h_{\phi}(\tau) \le 1 	\,.
\end{align}
For a meaningful violation of Poisson statistics, $h_{\phi}(\tau)$ must be 
outside these bounds.

Also, $h_{\phi}(\tau)$ is a measure of non-Gaussian fluctuations, here of 
third-order in the field. Resonance fluorescence is a particularly strong 
case of non-Gaussian noise by being a highly nonlinear stationary 
nonequilibrium process \cite{CaRG16,GCRH17,SaCa21,XGJM15}, thanks 
also to its small Hilbert space. 

\subsection{Fluctuations Spectra} 	
Since quadrature fluctuations, such as squeezing, are often studied 
in the frequency domain, we now define the spectrum of the 
amplitude-intensity correlations:  
\begin{eqnarray}
S_{\pi,\phi}(\omega) = 8 \gamma_1 \int_0^{\infty} d\tau \cos{(\omega \tau)} 
	\left[ h_{\pi,\phi}(\tau) -1 \right] 
\end{eqnarray}
which, following Eqs.~(\ref{eq:chd}) and (\ref{eq:chdsplit}), can be 
decomposed into terms of second- and third-order in the dipole fluctuations
\begin{eqnarray}
S_{\pi,\phi}^{(q)}(\omega) = 8 \gamma_1 \int_0^{\infty} d\tau 
	\cos{(\omega \tau)}  h_{\pi,\phi}^{(q)}(\tau) , 
\end{eqnarray}
where $q=2,3$, so that $S_{\pi,\phi}(\omega) = S_{\pi,\phi}^{(2)}(\omega) 
+S_{\pi,\phi}^{(3)}(\omega)$. 

As mentioned above, the AIC was devised initially to measure squeezing 
without the issue of imperfect detection efficiencies. Obviously, by definition, 
$h_{\pi,\phi} (\tau)$ and $S_{\pi,\phi}(\omega)$ are not measures of 
squeezing. They measure a third-order moment in the field's amplitude, 
while squeezing is a second-order one in its fluctuations. The so-called 
spectrum of squeezing is the one for $q=2$, with the advantage for the 
AIC of not depending on the efficiency of detection. Squeezing is 
signaled by frequency intervals where $S_{\pi,\phi}^{(2)}(\omega) <0$. 
As a further note, the full incoherent spectrum, Eq.~(\ref{eq:Sinc}), can be 
obtained by adding the squeezing spectra of both quadratures 
\cite{RiCa88},    
\begin{eqnarray}
S_{\pi}^{inc}(\omega) = \frac{1}{8\gamma_1}
	\left[ S_{\pi,0}^{(2)}(\omega) +S_{\pi,\pi/2}^{(2)}(\omega) \right] . 
\end{eqnarray}

\subsection{Results} 	
\begin{figure}[t]
\includegraphics[width=8.5cm,height=6.2cm]{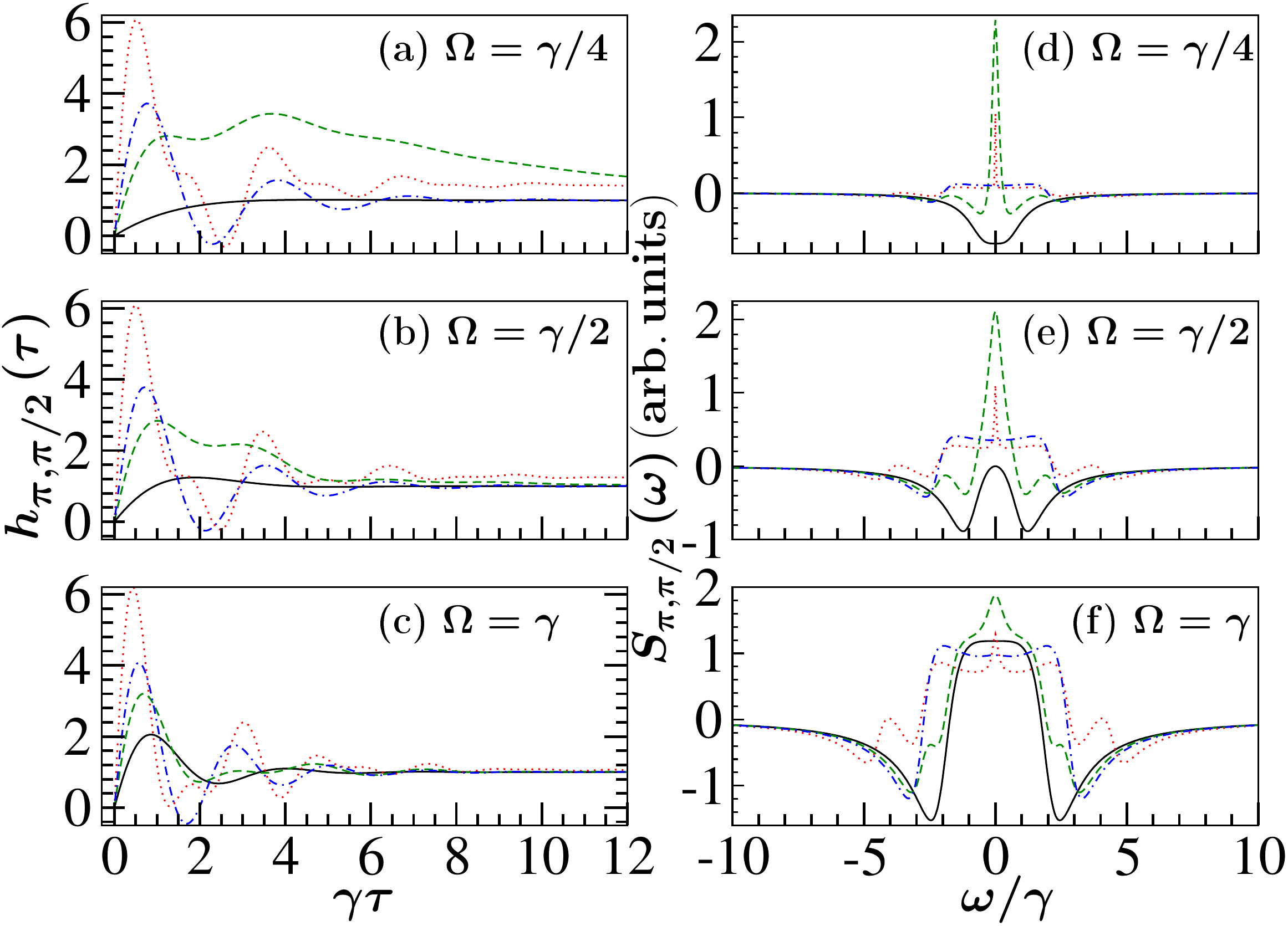} 
\caption{\label{fig:htauw} 
Amplitude-intensity correlations $h_{\pi,\phi}(\tau)$ (left panel) and 
corresponding spectra $S_{\pi,\phi} (\omega)$ (right 
panel) for the $\phi= \pi/2$ quadrature in the weak-moderate field limit. 
Parameters and line styles are the same as in Fig.~\ref{fig:g2}: 
$\Delta =\delta =0$ (solid-black); $\Delta = 2 \gamma$ and 
$\delta= -2\gamma$ (dots-red); $\Delta =-2 \gamma$ and 
$\delta= -2\gamma$ (dashed-green); $\Delta =-2 \gamma$ and 
$\delta= -4\gamma$ (dot-dashed-blue). }
\end{figure}
\begin{figure}[t]
\includegraphics[width=8.5cm,height=6.2cm]{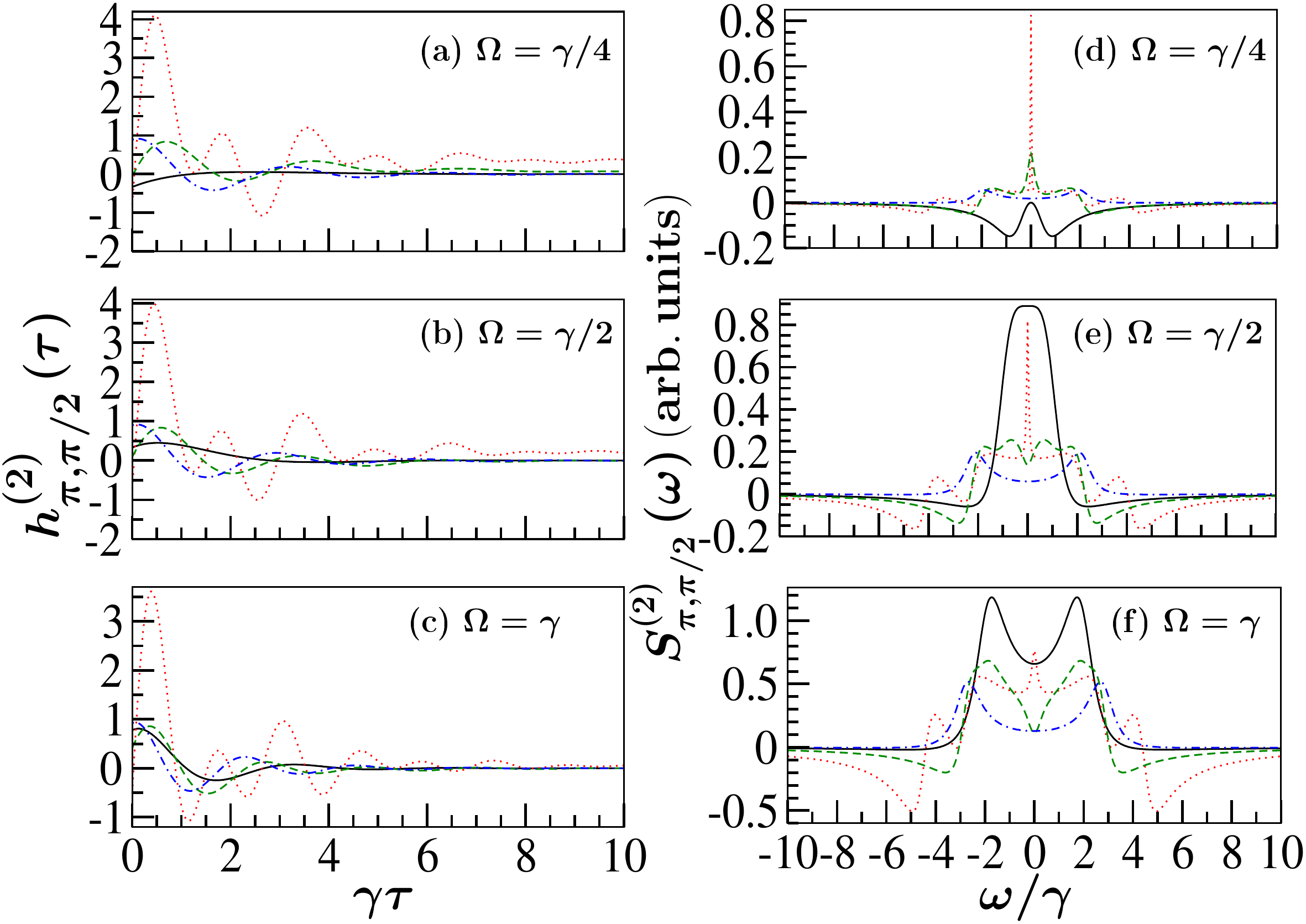} 
\caption{\label{fig:ht2w} 
Second-order component of the AIC, $h_{\pi,\phi}^{(2)}(\tau)$ (left panel), and corresponding (squeezing) spectra, $S_{\pi,\phi}^{(2)}(\omega)$ (right panel), 
with the same parameters and line styles as in Fig.~\ref{fig:htauw}.  }
\end{figure}

\begin{figure}[t]
\includegraphics[width=8.5cm,height=6.5cm]{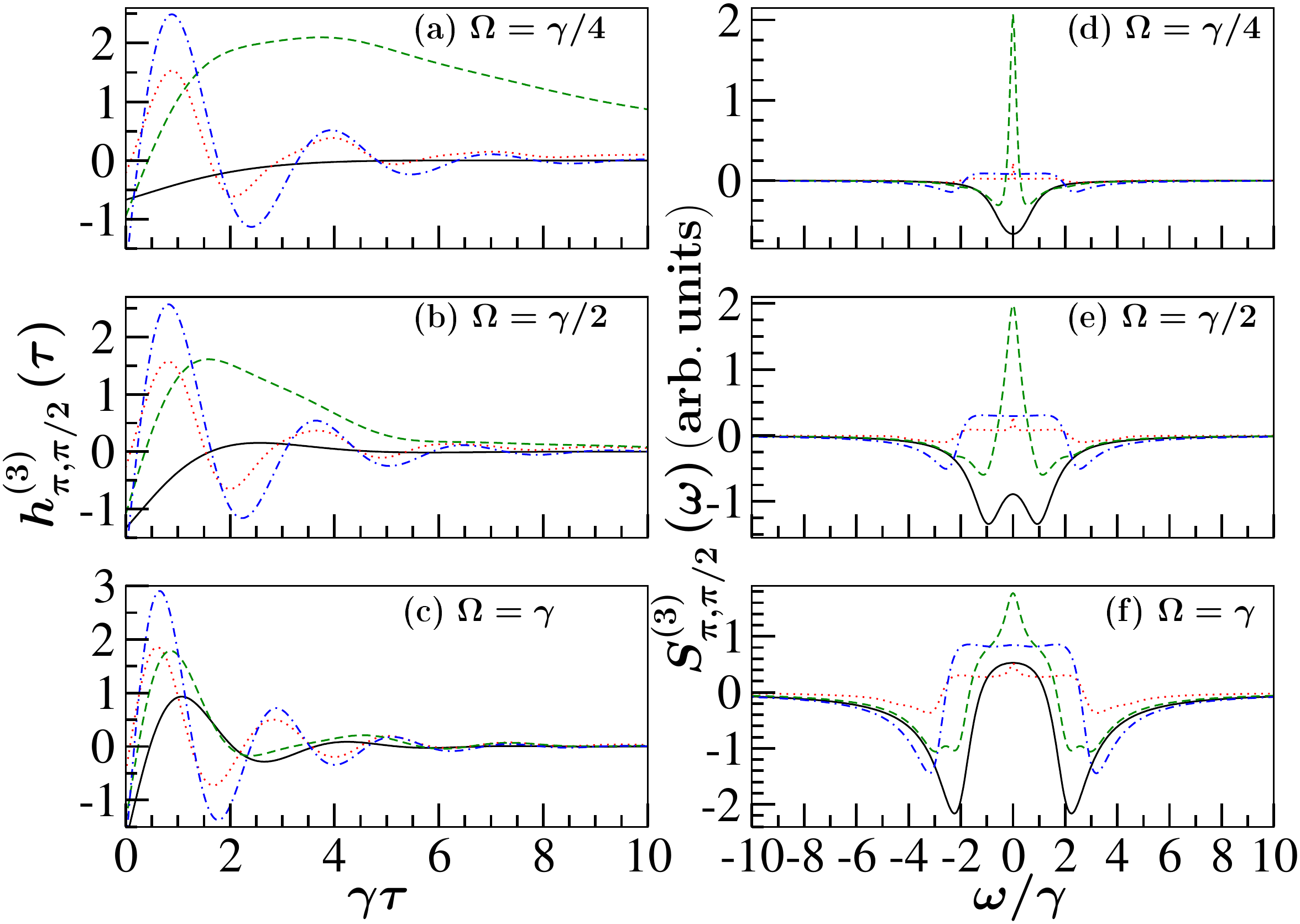} 
\caption{\label{fig:ht3w} 
Third-order component of the AIC, $h_{\pi,\phi}^{(3)}(\tau)$ (left panel), and 
corresponding spectra, $S_{\pi,\phi}^{(3)}(\omega)$ (right panel), with the 
same parameters and line styles as in Fig.~\ref{fig:htauw}. 
 }   
\end{figure}

 
We now show plots of the AICs and their spectra in 
Figs.~\ref{fig:htauw}-\ref{fig:ht3w} for the $\phi=\pi/2$ quadrature and the 
same sets of detunings $\Delta, \delta$ of Fig.~\ref{fig:pops}, and weak to 
moderate Rabi frequencies, $\gamma/4 < \Omega < \gamma$. Squeezing 
in the $\phi=0$ case will be considered only in the next Section. With the 
three parameters $\Omega$, $\Delta$, and $\delta$, there is a vast 
landscape of effects. 

We first notice a few general features in $h_{\pi,\pi/2}(\tau)$,  
Fig.~\ref{fig:htauw}. With increasing Rabi frequencies, detunings, and 
Zeeman splittings, we observe the clear breakdown of the classical 
inequalities besides the one at $\tau=0$. Correspondingly, in the spectra, 
the extrema get displaced and broadened. Now, we single out the case of 
nondegeneracy with small detuning on the $|1\rangle -|3\rangle$ transition 
but large on the $|2\rangle -|4\rangle$ one, $\Delta= -\delta =2\gamma$ 
(green-dashed line). For weak field, $\Omega= \gamma/4$, the AIC decays 
very slowly, with a corresponding very narrow spectral peak. The slow decay 
is also clearly visible in the photon correlation, Fig.~\ref{fig:g2}a. As we 
mentioned in Sect.~III regarding Fig.~\ref{fig:pops}b, state $|4\rangle$ ends 
up with a large portion of the steady-state population due to optical pumping; 
not quite a trapping state, so there is no electron shelving \textit{per se}, as 
argued in \cite{KiEK06b}. This effect is washed out for larger Rabi 
frequencies, allowing for faster recycling of the populations. To a lesser 
degree, slow decay and sharp peak occur for opposite signs of $\Delta$ 
and $\delta$. 

Splitting the AIC and spectra into components of second- and third-order 
fluctuations, Figs.~\ref{fig:ht2w} and \ref{fig:ht3w}, respectively, help us 
better understand the quadrature fluctuations. For $\Omega \ll \gamma$ 
and $\Delta=\delta=0,$ the squeezing spectrum is a negative central peak 
centered at $\omega=0$ (not shown). $\Omega = \gamma/4$ is already a 
strong enough driving that the squeezing gets displaced to sidebands and 
eventually gets lost (positive) for stronger fields. Finite laser and Zeeman 
detunings are detrimental to squeezing unless larger $\Omega$ is applied. 
For increasing Rabi frequencies, in $h_{\pi,\pi/2}^{(2)} (\tau)$ there is a 
reduction in amplitudes and nonclassicality except for the case of 
opposite signs of detuning and difference Zeeman splitting. Note that the 
sharp spectral peak in the latter case takes up most of the corresponding 
peak in Fig.~\ref{fig:htauw}. Increasing $\Omega$, the fluorescent 
emission becomes more incoherent, Eqs.(\ref{eq:meanIntensities}), and 
third-order fluctuations overcome the second-order ones. Also, comparing 
Figs.~\ref{fig:ht2w} and \ref{fig:ht3w} we see that $h_{\phi}^{(3)}(\tau)$ is 
mainly responsible for the breakdown of the classical bounds. Moreover, 
we see that the slow-decay--sharp-peak is mostly a third-order effect. 
\begin{figure}[t]
\includegraphics[width=8.5cm,height=9cm]{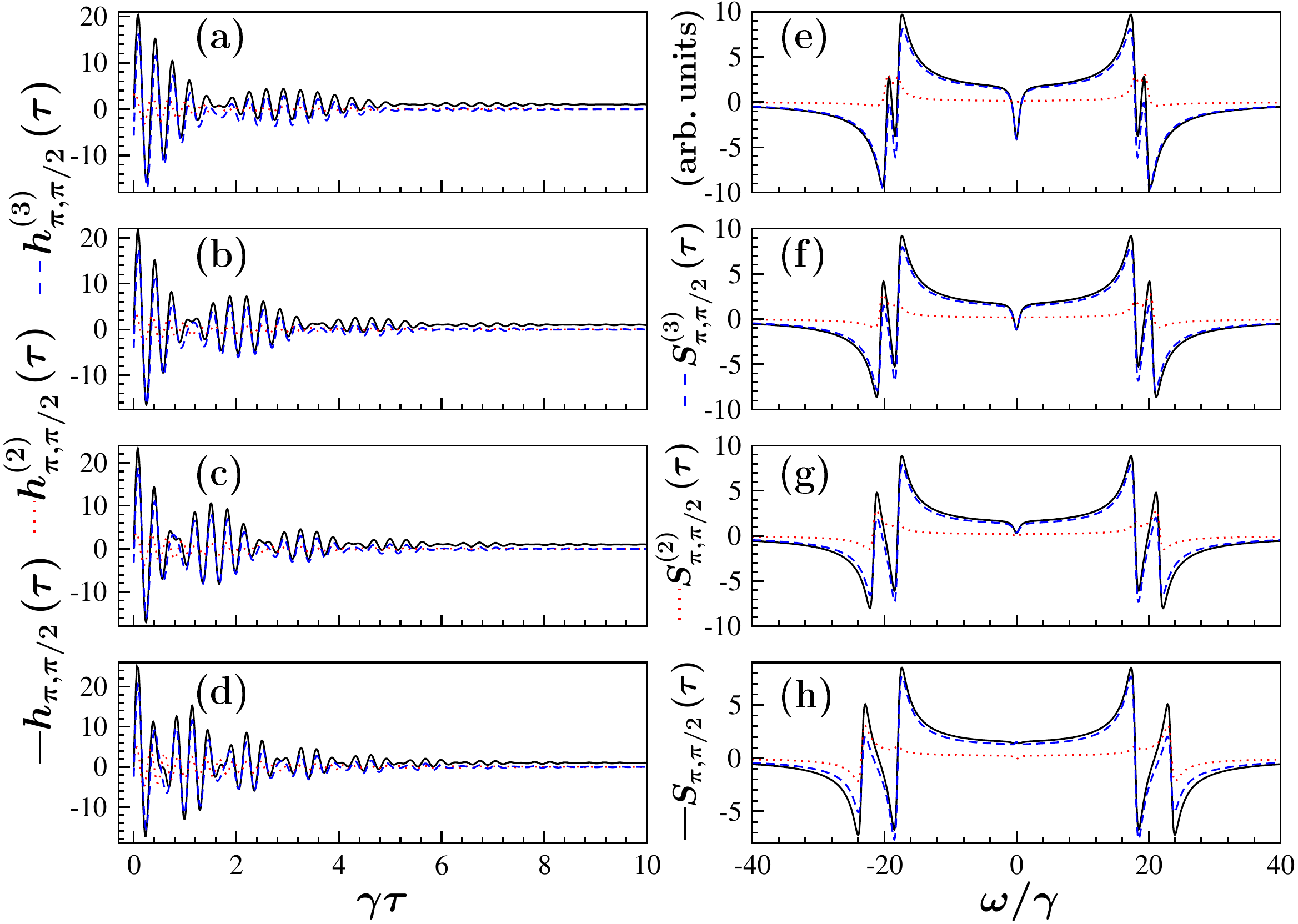} 
\caption{\label{fig:ht23s} 
Second-order (dots-red), third-order (dashed-blue), and total (solid-black)  
AICs (left panel) and corresponding spectra (right panel) for 
$\Omega = 9 \gamma$, $\Delta = 0$, and: (a,e) $\delta = -8\gamma$, 
(b,f) $\delta = -10\gamma$, (c,g) $\delta = -12\gamma$, (d,h) 
$\delta = -15\gamma$.   }   
\end{figure}

For very strong fields and large Zeeman splittings, 
$\Omega,|\delta| \gg \gamma$, Fig.~\ref{fig:ht23s}, the AIC shows beats as 
in the photon correlations. Unlike those in $g^{(2)}(\tau)$, the wavepackets 
here oscillate around $h(\tau) =1$. The spectral peaks are localized around 
the Rabi frequencies $\pm \Omega_1,\pm \Omega_2$ in a dispersive 
manner, revealing the wave character of the quadratures. 
 
Studies of the spectrum of squeezing for the $J=1/2 - J=1/2$ system were 
reported in \cite{TaXL09}, choosing to increase $\Omega_1,\Omega_2$ 
with large detunings, $\Delta > \Omega$, showing two close negative 
sidebands. In these conditions, beats would also occur but are not 
reported. The authors do, nonetheless, notice the interesting case where, 
for $\delta = 2\Delta$, $\Omega_1$ and $\Omega_2$ are equal, so the 
two sidebands merge into a single one. In this case, the beats would 
disappear.  

An odd-order function of the field amplitude opens the door naturally to 
probe for non-Gaussian fluctuations, the AIC reaching non-zero third-order 
moments, $\langle \Delta A_{ij} \Delta A_{kl} \Delta A_{mn} \rangle$. The 
latter are very small when the driving is weak, and the emission is mainly 
coherent but, as we saw above, they dominate for strong driving, and the 
emission is mostly incoherent. A few-level system, such as the 
$J=1/2 - J=1/2$ one,  is strongly nonlinear, so non-Gaussian fluctuations 
of its resonance fluorescence are ubiquitous. With the AIC, unlike moments 
in photon counting, which are positive-definite, the fluctuations can also 
take negative values and break classical bounds of quadratures in both time 
and frequency domains \cite{CaRG16,SaCa21}. Quantum beats, a major 
signature of interference, are found to be also non-Gaussian in resonance 
fluorescence, as shown particularly in Fig.~\ref{fig:ht23s}.    

\section{Variance}  	\label{sec:variance}
The variance is a measure of the total noise in a quadrature; it is defined as 
\begin{eqnarray}
V_{\phi} &=& \langle :(\Delta E_{\phi})^2: \rangle 	
	= \mathrm{Re} \left[ e^{-i\phi} 
	\langle \Delta \hat{E}^-  \Delta \hat{E}_{\phi}  \rangle_{st}  \right] , 
\end{eqnarray}
and is related to the spectrum of squeezing  as 
\begin{eqnarray}
V_{\phi} &=& \frac{1}{4\pi \gamma \eta} 
	\int_{-\infty}^{\infty} d\omega S_{\phi}^{(2)} (\omega) .
\end{eqnarray}
where $\eta$ is the detector efficiency. The maximum value of $V_{\phi}$ is 
1/4, obtained when there is very strong driving, when almost all the emitted 
light is incoherent. Negative values of the variance are a signature of 
squeezing but, unlike the quadrature spectra, the squeezing is the total one 
in the field, independent of frequency.  

For the $\pi$ transitions we have
\begin{subequations}
\begin{eqnarray}
V_{\pi,\phi} &=& \frac{ f_{\pi}^2(r) }{2} \mathrm{Re} 
	\left[ -( \alpha_{13} -\alpha_{24})^2 e^{-2i\phi} \right. \nonumber \\
	&& \left. +( \alpha_{11} +\alpha_{22} -|\alpha_{13} 
	 -\alpha_{24}|^2 )  \right]  	  \\
&=& 	\frac{ f_{\pi}^2(r) }{2} \frac{\Omega^2}{D} \left[ 1 
	- \frac{ [(2\Delta -\delta)\cos{\phi} +\gamma\sin{\phi}]^2}{2D}  \right] . 
	\nonumber \\ 
\end{eqnarray}
\end{subequations}
For $\phi=\pi/2$ and $\phi=0$ we have 
\begin{subequations}
\begin{eqnarray} 	
V_{\pi,\pi/2} &=& \frac{ f_{\pi}^2(r) }{2} \frac{\Omega^2}{D} 
	\left[ 1 - \frac{\gamma^2}{2D}  \right] , 	\\
V_{\pi,0} &=& \frac{ f_{\pi}^2(r) }{2} \frac{\Omega^2}{D} 
	\left[ 1 - \frac{(2\Delta -\delta)^2}{2D}  \right] ,
\end{eqnarray}
\end{subequations}
respectively, where $D$ is given by Eq.~(\ref{eq:denominator}). 

\begin{figure}[t]
\includegraphics[width=8.55cm,height=8cm]{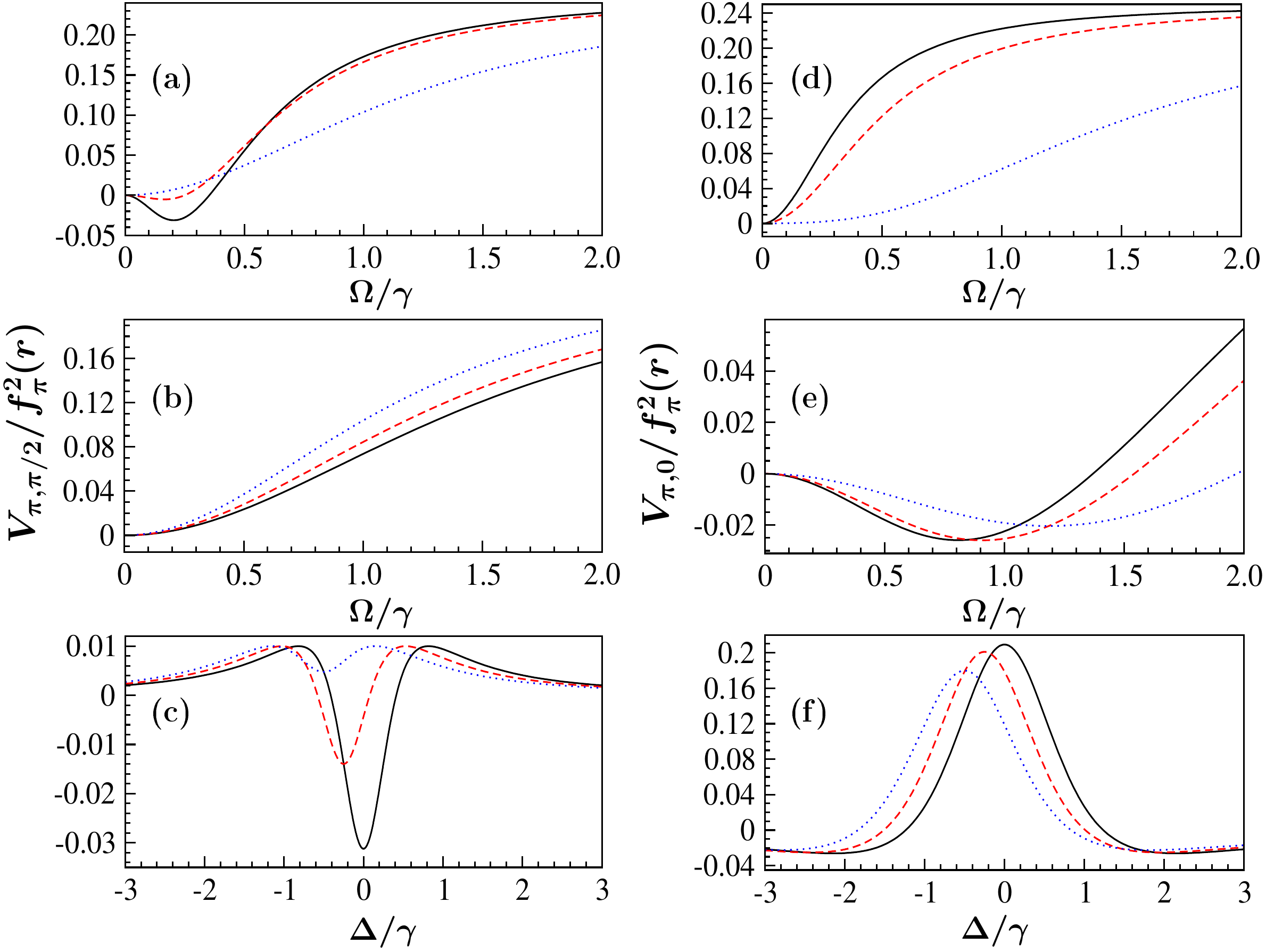} 
\caption{\label{fig:variance} 
Variance of the quadratures of the fluorescence of the $\pi$ transitions:  
left panel for $\phi=\pi/2$ and right panel for $\phi=0$. (a,b,d,e) as a 
function of Rabi frequency and (c,f) as a function of laser detuning. 
In all cases, $\delta=0$ is given by a solid-black line, and 
$\delta=-0.5\gamma$ by a dashed-red line; the dotted-blue line is 
$\delta=-2\gamma$ in (a,b,d,e) and $\delta=-\gamma$ in (c,f). Additionally, 
(a) $\Delta=0$,  (b) $\Delta=-2 \gamma$, (c) $\Omega=0.2 \gamma$, 
(d) $\Delta=-2\gamma$, 
(e) $\Delta=2\gamma$, (f) $\Omega=0.8 \gamma$.  }
\end{figure}

In Fig.~\ref{fig:variance} we plot the variances of the out-of-phase 
$\phi=\pi/2$ (left panel) and in-phase $\phi=0$ (right panel) quadratures. 
The interplay of parameters is a complex one, but we mostly use the 
ones of previous figures. For $\phi=\pi/2$, as usual in single-atom 
resonance fluorescence systems, squeezing is restricted to small Rabi 
frequencies; increasing laser and Zeeman detunings are detrimental to 
squeezing. In contrast, for $\phi=0$ nonzero laser or Zeeman detunings 
are necessary to produce squeezing, with a strong dependence on their 
sign: on-resonance (not shown), there is no squeezing, a result also known 
for a two-level atom; in Fig.~\ref{fig:variance}(d) the laser is tuned below 
the $|1 \rangle - |3 \rangle$ transition, $\Delta = - 2\gamma$, and there is 
no squeezing (positive variance), but the variance is reduced for large 
$\delta$; in Fig.~\ref{fig:variance}(e) the laser is tuned above that 
transition, $\Delta = 2\gamma$, and there is squeezing for not very large 
Rabi frequencies. Large values of $\delta$ tend to reduce the variance, 
be it positive or negative. 

\subsection{Out-of-phase quadrature} 	
We now discuss a complementary view of the variance. For $\phi=\pi/2$ we 
can identify the Rabi frequency interval within which squeezing takes place, 
\begin{eqnarray}
0 < \Omega < \frac{1}{2} \sqrt{\gamma^2/2 - \delta^2/2 -2(\Delta -\delta/2)^2} ,
\end{eqnarray} 
and the Rabi frequency for maximum squeezing is 
\begin{eqnarray}
\tilde{\Omega}_{\pi/2} = \frac{1}{2} \sqrt{ 
	\frac{ \gamma^4/2 -2[(\delta -\Delta)^2 +\Delta^2]^2 }
	{ 3\gamma^2 + 2[(\delta -\Delta)^2 +\Delta^2]^2} } .
\end{eqnarray} 
Thus, the variance at $\tilde{\Omega}_{\pi/2}$ is 
\begin{subequations}
\begin{eqnarray}
V_{\pi,\pi/2}^{(\tilde{\Omega}_{\pi/2})} (\Delta=0,\delta) 
	&=& \frac{ f_{\pi}^2(r)}{16} 
	\frac{ (\gamma^4/2 -2\delta^4)(\delta^2 -\gamma^2) }
	{ \gamma^2(\gamma^2 +2\delta^2)(\delta^2 +\gamma^2) } , 
	\nonumber \\
\end{eqnarray}
for $\Delta=0$ and $|\delta/\gamma| < 1/\sqrt{2}$;  
\begin{eqnarray}
V_{\pi,\pi/2}^{(\tilde{\Omega}_{\pi/2})}  (\Delta, \delta=0) 
	&=& \frac{ f_{\pi}^2(r)}{16} 
	\frac{ (\gamma^4/2 -8\Delta^4)(4\Delta^2 -\gamma^2) }
	{ \gamma^2(\gamma^2 +4\Delta^2)^2 } , \nonumber \\
\end{eqnarray}
for $\delta=0$ and $|\Delta/\gamma| < 1/\sqrt{2}$; and the maximum 
total squeezing is obtained at $\Delta =\delta =0$,  
\begin{eqnarray}
V_{\pi,\pi/2}^{(\tilde{\Omega}_{\pi/2})} (0,0) &=& - \frac{ f_{\pi}^2(r)}{32} , 
	\qquad \tilde{\Omega}_{\pi/2} = \frac{\gamma}{2\sqrt{6}} . 
\end{eqnarray}
\end{subequations}

For $\phi=\pi/2$  squeezing is limited to elliptical regions of weak driving 
and small detunings $\Delta$ and $\delta$:   
\begin{subequations}
\begin{eqnarray}
2\delta^2 +8\Omega^2 &<& \gamma^2 ,   \qquad \Delta=0 , \\ 
4\Delta^2 +8\Omega^2 &<& \gamma^2 ,   \qquad \delta=0 .
\end{eqnarray}
\end{subequations}

These results indicate that the conditions for squeezing in the resonance 
fluorescence of the $J=1/2 - J=1/2$ system are more stringent than that 
for a two-level atom \cite{CaRG16}: it is limited to smaller Rabi frequencies 
and the minimum variance is also smaller. Hence, there is no net 
squeezing in the regime of quantum beats seen in the previous Sections.

\subsection{In-phase quadrature} 	
For $\phi=0$, squeezing is obtained in the Rabi frequency interval, for 
$\delta=0$, 
\begin{eqnarray}
0 < \Omega < \frac{1}{\sqrt{2}} \sqrt{ \Delta^2 -\gamma^2/4} , 
	\qquad |\Delta| > \gamma/2 , 
\end{eqnarray} 
with maximum squeezing at the Rabi frequency 
\begin{eqnarray}
\tilde{\Omega}_0 = \frac{1}{2\sqrt{2}} \sqrt{ 
	\frac{ 16 \Delta^2 -\gamma^2 }{ 12 \Delta^2 +\gamma^2 }   } ,
\end{eqnarray} 
requiring finite detuning from both $\pi$ transitions ($\Delta \neq 0$) 
and stronger driving, $\Omega \sim \gamma$ [see 
Fig.~\ref{fig:variance}(d)-14(f)]. 

Thus, the variance at $\tilde{\Omega}_0$ is 
\begin{eqnarray}
V_{\pi,0}^{(\tilde{\Omega}_0)}  (\delta) &=& -\frac{ f_{\pi}^2(r)}{128}  
	\frac{ 4\Delta^2 -\gamma^2 }{ \Delta^2 (4\Delta^2 +\gamma^2) } , 
	\quad |\Delta| \geq \gamma/2 .
\end{eqnarray}
This expression gets the asymptotic value
\begin{eqnarray}
\lim_{\Delta \to \infty} V_{\pi,0}^{(\tilde{\Omega}_0)}  
	&=& -\frac{ f_{\pi}^2(r)}{32}  ,
\end{eqnarray}
which is the same as that for the $\pi/2$ quadrature. The region for 
squeezing obeys the relation 
\begin{eqnarray}
4\Delta^2 -8\Omega^2 < \gamma^2 .  
\end{eqnarray}
So, to obtain squeezing in this quadrature, it is necessary to have 
detunings $\Delta > \gamma/4$ for any Rabi frequency.

\section{Conclusions}   
We have studied interference effects on the resonance fluorescence of the 
$\pi$ transitions in a $J=1/2 - J=1/2$ angular momentum atomic system 
driven by a linearly polarized laser field and a magnetic field to lift the level 
degeneracies. These fields make the transition frequencies unequal, so the 
$\pi$ transitions evolve with different generalized Rabi frequencies, which 
we observed first in the time-dependent populations of the excited states  
and whose interference produces quantum beats. When the atom is subject 
to large laser and magnetic fields, the beats have well-defined modulation of 
the fast oscillations. We studied beats in the total intensity and two-time 
correlations. It is generally found that the main contributions to these 
observables come from the interference of the individual $\pi$ transitions 
density probabilities, while those due to vacuum-induced coherence, which 
link both transition amplitudes, have a lesser role. This is because the upper 
levels are very separated due to the large ac and Zeeman splittings.  
   
Nonclassical features of the fluorescence light are shown in photon 
correlations, such as antibunching, $g^{(2)}(0) =0$, and in 
amplitude-intensity correlations, with squeezing and violation of classical 
inequalities, Eqs.~(\ref{eq:ineq}, \ref{eq:ineq_cs}). 
In resonance fluorescence, interference is generally detrimental to 
squeezing, even at moderate laser and magnetic fields. In the regime of 
well-defined beats, there is squeezing near the effective Rabi frequencies 
but not in the total noise, as the variance confirms. Also, in this regime, 
third-order fluctuations in the field quadrature amplitude dominate over the 
second-order ones and break classical inequalities, making the fluctuations 
strongly non-Gaussian, making the beats and their spectra outstanding 
features. Resonance fluorescence, being a highly nonequilibrium and 
nonlinear stationary process, is a useful laboratory for the study of 
nonclassical and non-Gaussian fluctuations. The AIC, with its conditional 
character (based on recording only pairs of events) and third-order 
moments in the field amplitudes, provides a practical and proven alternative 
to more recent work based on time-gating photodetection \cite{Furusawa23} 
and higher odd-order moments in photoncounting \cite{Clerk20,Sifft23}.
  
So far, few quantum interference experiments have been performed on the 
$J=1/2 - J'=1/2$ system, but we believe experiments to observe our results 
are within reach. Certainly, single-atom resonance fluorescence is bound to 
both photon collection and detection inefficiencies. Conditional measurements 
such as photon-photon and photon-quadrature correlations solve this issue; 
also, through (inverse) Fourier transform of the spectrum might be an efficient 
way to observe beats. However, both laser-atom and Zeeman detunings 
should not be too large since they imply reduced fluorescence rates 
$\gamma_1 (\alpha_{11}+\alpha_{22})$, see Eq.(\ref{eq:BlochSteady1}), 
which may be detrimental in measurements. Also, a large difference Zeeman 
splitting means that the upper levels would be very separated, diminishing 
the vacuum-induced coherence. The beats would be better observed if 
$\Delta \leq \gamma$ and $\delta$ of just several $\gamma$ in the strong 
field regime, $\Omega \gg \gamma$. In this regime, the beats are due 
basically to interference among the incoherently coupled waves of the two 
$\pi$ transitions, but bound by the $\sigma$ transitions. Finally, we have 
used \textit{interference}  to mean not only the presence of cross-terms as 
in Young-type interference but also the superposition of coupled transitions 
as discussed in this paper.

\section{Acknowledgments.} 
The authors thank Dr. Ricardo Rom\'an-Ancheyta for continued interest 
in this work and Dr. Ir\'an Ramos-Prieto for useful comments at an early 
stage of the project. ADAV thanks CONACYT, Mexico, for the support of 
scholarship No. 804318. 

ORCID numbers:
H\'ector M. Castro-Beltr\'an https://orcid.org/0000-0002-3400-7652, 
Octavio de los Santos-S\'anchez https://orcid.org/0000-0002-4316-0114,
Luis Guti\'errez https://orcid.org/0000-0002-5144-4782, 
Alan D. Alcantar-Vidal https://orcid.org/0000-0003-3849-7844.

\appendix
\section{Time-Dependent Solutions of the Matrix Equations and Spectra 
by Formal Integration of Correlations \label{ap:matrixSol}} 
The two-time photon correlations under study have the general form 
$\langle \mathbf{W}(\tau) \rangle 
=  \langle O_1 (0) \mathbf{R}(\tau) O_2(0) \rangle$, where $\mathbf{R}$ 
is the Bloch vector and $O_{1,2}$ are system operators. The same applies 
to correlations of fluctuation operators $\Delta \mathbf{R}, \,\Delta O_{1,2}$. 
Using the quantum regression formula \cite{Carm02}, the correlations obey 
the equation 
\begin{eqnarray} 	\label{eq:EqQRF}
\langle \dot{\mathbf{W}} (\tau) \rangle 
	= \mathbf{M} \langle \mathbf{W}(\tau) \rangle  ,	
\end{eqnarray}
which has the formal solution
\begin{eqnarray} 	\label{eq:SolQRF}
	\langle \mathbf{W} (\tau) \rangle 
		= e^{ \mathbf{M} \tau}  \langle \mathbf{W}(0) \rangle  ,
\end{eqnarray}
where $\mathbf{M}$ is given by 
\begin{widetext}
\begin{eqnarray} 	\label{eq:matrixM}
\mathbf{M} = \left( \begin{array}{cccccccc} 
-\gamma &-i \Omega &0 &0 & i \Omega &0 &0 &0 		\\ 
-i \Omega &-\left(\frac{\gamma}{2} +i\Delta \right) &0 &0 &0 &i \Omega &0 &0 \\ 
0 &0 & -\gamma &i \Omega &0 &0 & -i \Omega &0 	\\ 
0 &0 &i \Omega &-\left(\frac{\gamma}{2} +i(\Delta-\delta) \right) 
	&0 &0 &0 &-i \Omega 	\\
i \Omega &0 &0 &0 & -\left(\frac{\gamma}{2} -i\Delta \right) &-i \Omega &0 &0 \\ 
\gamma_1 &i \Omega &\gamma_{\sigma} &0 &-i \Omega &0 &0 &0 	\\ 
0 &0 &-i \Omega &0 &0 &0 &-\left(\frac{\gamma}{2} -i(\Delta-\delta) \right) 
	&i \Omega \\ 
\gamma_{\sigma} &0 &\gamma_2 &-i \Omega &0 &0 &i \Omega &0 
	\end{array} \right) 	.
\end{eqnarray}
\end{widetext}

Also, the spectra of stationary systems can be evaluated more effectively 
using the above formal approach. Be 
$g(\tau) = \langle \mathbf{W}(\tau) \rangle$.   
Then, a spectrum is calculated as 
\begin{eqnarray} 	\label{eq:genspectrum}
S(\omega) &\propto& \int_0^{\infty}  \cos{\omega \tau} \,g(\tau) \, d\tau 
	= \int_0^{\infty} \cos{\omega \tau} \,e^{ \mathbf{M} \tau } g(0)  \, d\tau 
	\nonumber \\ 
&=& \mathrm{Re}  \int_0^{\infty} 
	e^{ - (i \omega \mathbf{1} - \mathbf{M} ) \tau } g(0)  \, d\tau \nonumber \\ 
&=& \mathrm{Re}  \left[ (i \omega \mathbf{1} -\mathbf{M})^{-1} g(0) \right] ,
\end{eqnarray}
where $\mathbf{1}$ is the identity matrix. For example, the incoherent 
spectrum requires calculations of the type 
\begin{eqnarray} 	
S^{inc}(\omega) &=&  \mathrm{Re} \int_0^{\infty} d\tau 
	e^{-i \omega \tau} e^{\mathbf{M} \tau} 
	\langle \Delta A_{ij}(0) \Delta A_{kl}(0) \rangle_{st} \nonumber \\ 
	&=&  \mathrm{Re} \left[ (\mathbf{M} - i\omega \mathbf{1})^{-1} 
		\langle \Delta A_{ij}(0) \Delta A_{kl}(0) \rangle_{st} \right] . \qquad 
\end{eqnarray}

For the initial conditions of the correlations we use the following operator 
products and correlations in compact form: 
\begin{subequations}
\begin{eqnarray} 	
A_{kl} A_{mn} &=& A_{kn} \delta_{lm} \,, 	\\
\langle A_{kl} A_{mn} \rangle &=& \alpha_{kn} \delta_{lm} , \\
A_{ij} A_{kl} A_{mn} &=& A_{in} \delta_{jk} \delta_{lm} , 	\\
\langle A_{ij} A_{kl} A_{mn} \rangle &=& \alpha_{in} \delta_{jk} \delta_{lm}  .  
\end{eqnarray}
\end{subequations}
Hence, the relevant initial conditions are:
\begin{subequations}
\begin{eqnarray} 	\label{eq:AjkR}
\langle A_{13} \mathbf{R} \rangle 
	&=& \left( 0, 0, 0, 0, \alpha_{11}, \alpha_{13},  0, 0   \right)^T , 	\\
\langle A_{24} \mathbf{R} \rangle 
	&=& \left( 0, 0, 0, 0, 0, 0, \alpha_{22}, \alpha_{24}  \right)^T , 	\\
\langle A_{13} \mathbf{R} A_{31} \rangle 
	&=& \left( 0, 0, 0, 0, 0, \alpha_{11}, 0, 0  \right)^T , \\
\langle A_{24} \mathbf{R} A_{42} \rangle 
	&=& \left( 0, 0, 0, 0, 0, 0, 0, \alpha_{22}  \right)^T , 	\\ 
\langle A_{13} \mathbf{R} A_{42} \rangle 
	&=& \langle A_{24} \mathbf{R} A_{31} \rangle = 0 , 	\label{eq:AjkRe}
\end{eqnarray}
\end{subequations}
where $\mathbf{R} = \left( A_{11}, A_{13}, A_{22}, A_{24}, 	A_{31}, 
A_{33}, A_{42}, A_{44} \right)^T$ is the Bloch vector. For correlations with 
fluctuation operator products,  $\Delta A_{ij} = A_{ij} - \alpha_{ij}$, we have  
\begin{eqnarray}
\langle \Delta A_{kl} \Delta A_{mn} \rangle 
	&=& \alpha_{kn} \delta_{lm} -\alpha_{kl} \alpha_{mn} , 	\\ 
\langle \Delta A_{ij} \Delta A_{kl} \Delta A_{mn} \rangle 
  &=& \alpha_{in} \delta_{lm} \delta_{jk}  -\alpha_{il} \alpha_{mn} \delta_{jk} 
  	\nonumber \\
  	  && - \alpha_{in} \alpha_{kl} \delta_{jm} 
	  	- \alpha_{ij} \alpha_{kn} \delta_{lm} 	\nonumber \\
	 && +2 \alpha_{ij} \alpha_{kl} \alpha_{mn}	.
\end{eqnarray}

Now, recalling that 
$\alpha_{12} = \alpha_{14} = \alpha_{23} = \alpha_{34} = 0$, we write the 
detailed initial conditions of the correlations (Bloch equations and quantum 
regression formula): 
\begin{widetext}
\begin{subequations}
\begin{eqnarray} 	
\langle \Delta A_{13} \Delta \mathbf{R} \rangle 
	&=& \left( -\alpha_{13} \alpha_{11},  \, -\alpha_{13}^2,  
	     \, -\alpha_{13} \alpha_{22},  \, -\alpha_{13} \alpha_{24} ,
  \, \alpha_{11} -|\alpha_{13}|^2, \, \alpha_{13} -\alpha_{13} \alpha_{33},  
	    \, -\alpha_{13} \alpha_{42},  \, -\alpha_{13} \alpha_{44}  \right)^T \,, 
	    \qquad \\
\langle \Delta A_{24} \Delta \mathbf{R} \rangle 
	&=& \left( -\alpha_{24} \alpha_{11},  \,-\alpha_{24} \alpha_{13}, 
	\, -\alpha_{24} \alpha_{22}, \, -\alpha_{24}^2, \,  -\alpha_{24} \alpha_{31}, 
	     \, -\alpha_{24} \alpha_{33}, \, \alpha_{22} -|\alpha_{24}|^2,  
	    \,  \alpha_{24} -\alpha_{24} \alpha_{44} 			\right)^T , 	 \qquad
\end{eqnarray}
\begin{eqnarray} 	
\langle \Delta A_{13} \Delta \mathbf{R} \Delta A_{31} \rangle 
	&=& \left(  2|\alpha_{13}|^2 \alpha_{11} -\alpha_{11}^2, 
	\, 2|\alpha_{13}|^2 \alpha_{13} -2\alpha_{11} \alpha_{13},  
		\right. \nonumber \\ 
	&& \left. 2|\alpha_{13}|^2 \alpha_{22} -\alpha_{11} \alpha_{22},   
	\,2|\alpha_{13}|^2 \alpha_{24} -\alpha_{11} \alpha_{24} ,   
		\right. \nonumber \\ 
	&&\left. 2|\alpha_{13}|^2 \alpha_{31} -2\alpha_{11} \alpha_{31}, 
	\,2|\alpha_{13}|^2 \alpha_{33} +\alpha_{11} -2 |\alpha_{13}|^2 
		-\alpha_{11} \alpha_{33},   
			\right. \nonumber \\ 
	&& \left. 2|\alpha_{13}|^2 \alpha_{42} -2\alpha_{11} \alpha_{42}, 
	\,2|\alpha_{13}|^2 \alpha_{44} -\alpha_{11} \alpha_{44}   \right)^T . 
\end{eqnarray}
\begin{eqnarray} 	
\langle \Delta A_{24} \Delta \mathbf{R} \Delta A_{42} \rangle 
	&=& \left(  2|\alpha_{24}|^2 \alpha_{11} -\alpha_{11} \alpha_{22} ,  
	\,2|\alpha_{24}|^2 \alpha_{13} -\alpha_{22} \alpha_{13}, 
		\right. \nonumber \\ 
	&& \left. 2|\alpha_{24}|^2 \alpha_{22} -\alpha_{22}^2,   
	\, 2|\alpha_{24}|^2 \alpha_{24} -2\alpha_{22} \alpha_{24} ,   
		\right. \nonumber \\ 
	&& \left. 2|\alpha_{24}|^2 \alpha_{31} -\alpha_{22} \alpha_{31},    
	\, 2|\alpha_{24}|^2 \alpha_{33} -\alpha_{22} \alpha_{33} ,     
		\right. \nonumber \\ 
	&& \left. 2|\alpha_{24}|^2 \alpha_{42} -2\alpha_{22} \alpha_{42} ,   
	\, 2|\alpha_{24}|^2 \alpha_{44} +\alpha_{22} -2|\alpha_{24}|^2
		-\alpha_{22} \alpha_{44} 	\right)^T .
\end{eqnarray}
\begin{eqnarray} 	
\langle \Delta A_{13} \Delta \mathbf{R} \Delta A_{42} \rangle 
	&=&  \left( 2\alpha_{13} \alpha_{11} \alpha_{42}, 
	\,2\alpha_{13}^2 \alpha_{42},  
	\, 2\alpha_{13} \alpha_{22} \alpha_{42}, 
	\, (2 |\alpha_{24}|^2 -\alpha_{22}) \alpha_{13} ,
		\right. \nonumber  	\\ 
	&& \left. (2|\alpha_{13}|^2 -\alpha_{11}) \alpha_{42},  
	\, (2\alpha_{13} \alpha_{33} -\alpha_{13}) \alpha_{42},  
		\, 2\alpha_{13} \alpha_{42}^2,  
	\, (2\alpha_{13} \alpha_{44} -\alpha_{13}) \alpha_{42} \right)^T , \\
\langle \Delta A_{24} \Delta \mathbf{R} \Delta A_{31} \rangle 
	&=& \left( 2\alpha_{24} \alpha_{11} \alpha_{31},  
	\, (2 |\alpha_{13}|^2 -\alpha_{11}) \alpha_{24},  
	\, 2\alpha_{24} \alpha_{22} \alpha_{31},  \, 
		2\alpha_{24}^2 \alpha_{31} ,\right. \nonumber  	\\ 
	&& \left. 2\alpha_{24} \alpha_{31}^2,   
	\, (2\alpha_{24} \alpha_{33} -\alpha_{24}) \alpha_{31},  
		 \, (2|\alpha_{24}|^2 -\alpha_{22}) \alpha_{31}, 
	\, (2\alpha_{24} \alpha_{44} -\alpha_{24}) \alpha_{31}  \right)^T .
\end{eqnarray}
\end{subequations}
\end{widetext}


\section{Condition for Optimal Appearance of Beats in the Intensity 
\label{beats_intensity}} 
We consider a simplified, unitary, model to estimate the optimal initial 
population of the ground states to make well-formed beats. First, we 
diagonalize the Hamiltonian Eq.~(\ref{eq:Hamiltonian}). The eigenvalues 
and eigenstates are   
\begin{subequations}
\begin{eqnarray}  	
\mathcal{E}_1^{\pm} 
	&=& - \frac{\Delta}{2}  \pm \frac{1}{2} \sqrt{ 4\Omega^2 +\Delta^2 } , \\ 
\mathcal{E}_2^{\pm} 
	&=& B_{\ell} + \frac{ \delta -\Delta}{2}  
		\pm \frac{1}{2} \sqrt{ 4\Omega^2 +(\delta -\Delta)^2 }    , 
\end{eqnarray}
\end{subequations}
and 
%
\begin{eqnarray} 	
|u_1\rangle &=& \sin{\Theta_1} |1\rangle +\cos{\Theta_1} |3\rangle , 
	\nonumber\\ 
|u_2\rangle &=& -\cos{\Theta_1} |1\rangle +\sin{\Theta_1} |3\rangle , 
	\nonumber\\
|u_3\rangle &=& \sin{\Theta_2} |2\rangle +\cos{\Theta_2} |4\rangle , 
	\nonumber\\
|u_4\rangle &=& -\cos{\Theta_2} |2\rangle +\sin{\Theta_2} |4\rangle ,
\end{eqnarray}
respectively, where 
\begin{eqnarray*}  	
\sin{\Theta_1} &=& \frac{2\Omega}
	{\sqrt{ \left( \Delta +\sqrt{\Delta^2 +4\Omega^2 } \right)^2 +4\Omega^2 }} , \\
\cos{\Theta_1} &=& \frac{ \Delta +\sqrt{\Delta^2 +4\Omega^2} }
	{\sqrt{ \left( \Delta +\sqrt{\Delta^2 +4\Omega^2 } \right)^2 +4\Omega^2 }} , 
	\nonumber 
\end{eqnarray*}
\begin{eqnarray}  	
\sin{\Theta_2} &=& \frac{2\Omega}
	{\sqrt{ \left( (\delta-\Delta) +\sqrt{(\delta -\Delta)^2 +4\Omega^2 } \right)^2 		+4\Omega^2 }} , \nonumber \\
\cos{\Theta_2} &=& \frac{ (\delta -\Delta) +\sqrt{(\delta-\Delta)^2 +4\Omega^2} }
	{\sqrt{ \left( (\delta-\Delta) +\sqrt{(\delta -\Delta)^2 +4\Omega^2 } \right)^2 		+4\Omega^2 }} . \nonumber \\
\end{eqnarray}

It is now straightforward to obtain the excited-state populations. If the initial 
state of the system is 
$\rho(0) = \langle A_{33} (0) \rangle |3\rangle \langle 3| 
+\langle A_{44}(0) \rangle |4\rangle \langle 4|$ 		we get 
\begin{subequations}
\begin{eqnarray} 	
\langle A_{33} (t) \rangle &=& \frac{1}{2} \langle A_{33} (0) \rangle 
	\sin^2{(2\Theta_1)} (1-\cos{(\Omega_1 t)}) , 	\qquad \\
\langle A_{44} (t) \rangle &=& \frac{1}{2} \langle A_{44} (0) \rangle 
	\sin^2{(2\Theta_2)} (1-\cos{(\Omega_2 t)}) , \qquad
\end{eqnarray}
\end{subequations}
and the intensity of the field is 
\begin{eqnarray} 	
\frac{I_{\pi} (\mathbf{r},t)}{ f_{\pi}^2(r)} 
	&=&  \langle A_{33} (0)  \rangle \sin^2{(2\Theta_1)} 
	+A_{44} (0)  \rangle \sin^2{(2\Theta_2)} 	\nonumber \\ 
&& - \langle A_{33} (0)  \rangle \sin^2{(2\Theta_1)} \cos{(\Omega_1 t)} 
	\nonumber \\ 
&& - \langle A_{44} (0)  \rangle \sin^2{(2\Theta_2)} \cos{(\Omega_2 t)}   .
\end{eqnarray} 

A necessary condition for the beating behavior to occur is that the initial 
ground-state populations are both nonvanishing in the nondegenerate 
case. Now, assuming the relation 
\begin{eqnarray} 	
\frac{ \langle A_{33} (0) \rangle }{ \langle A_{44} (0) \rangle } 
	= \frac{ \sin^2{(2\Theta_2)} }{ \sin^2{(2\Theta_1)} }
\end{eqnarray} 
is satisfied by choosing appropriate parameter values 
$(\Omega, \delta, \Delta)$ for given values of initial ground state populations 
we would get 
\begin{eqnarray} 	
I_{\pi} (\mathbf{r},t) &=& f_{\pi}^2(r)
	  \langle A_{33} (0)  \rangle \sin^2{(2\Theta_1)} 	\nonumber \\
   && \times  \left[ 1- \cos{(\Omega_{beat} t)} \cos{(\Omega_{av} t)} \right]	, 
\end{eqnarray} 
where $\Omega_{beat} = (\Omega_2 -\Omega_1)/2$ and 
$\Omega_{av} = (\Omega_2 +\Omega_1)/2$.


\end{document}